\documentclass[twocolumn,prb,aps,10pt]{revtex4-1}
\usepackage{graphicx,epsf}

\begin{document}

\title{Stochastic dynamics for a single vibrational mode in
molecular junctions}

\author{A. Nocera$^{1}$, C.A. Perroni$^{2}$, V. Marigliano Ramaglia$^{2}$ and V. Cataudella$^{2}$ }

\affiliation{$^{1}$ Dipartimento di Fisica ''E. Amaldi'',
Universita' di Roma Tre, Via della Vasca Navale 84, I-00146 Roma,
Italy\\
$^{2}$ CNR-SPIN and Universita' degli Studi di Napoli ''Federico II''\\
Complesso Universitario Monte Sant'Angelo, Via Cintia, I-80126
Napoli, Italy.}

\begin {abstract}
We propose a very accurate computational scheme for the dynamics
of a classical oscillator coupled to a molecular junction driven
by a finite bias, including the finite mass effect. We focus on
two minimal models for the molecular junction: Anderson-Holstein
(AH) and two-site Su-Schrieffer-Heeger (SSH) models. As concerns
the oscillator dynamics, we are able to recover a Langevin
equation confirming what found by other authors with different
approaches and assessing that quantum effects come from the
electronic subsystem only. Solving numerically the stochastic
equation, we study the position and velocity distribution
probabilities of the oscillator and the electronic transport
properties at arbitrary values of electron-oscillator interaction,
gate and bias voltages. The range of validity of the adiabatic
approximation is established in a systematic way by analyzing the
behaviour of the kinetic energy of the oscillator. Due to the
dynamical fluctuations, at intermediate bias voltages, the
velocity distributions deviate from a gaussian shape and the
average kinetic energy shows a non monotonic behaviour. In this
same regime of parameters, the dynamical effects favour the
conduction far from electronic resonances where small currents are
observed in the infinite mass approximation. These effects are
enhanced in the two-site SSH model due to the presence of the
intermolecular hopping $t$. Remarkably, for sufficiently large
hopping with respect to tunneling on the molecule, small
interaction strengths and at intermediate bias (non gaussian
regime), we point out a correspondence between the minima of the
kinetic energy and the maxima of the dynamical conductance.
\end {abstract}

\maketitle

\newpage

\section{Introduction}
In recent years it has become possible to fabricate electronic
devices where the effective element of a \emph{junction} is a
single molecule placed between two metallic (or semiconductor)
electrodes.$\cite{Tao,Reed,Reichert}$ Due to the small size of
molecules, the charging of the molecular bridge is often
accompanied by significant changes of the nuclear geometry,
indicating a strong coupling between electronic and nuclear (in
particular vibrational) degrees of freedom. For example, some
authors$\cite{Galperin05}$ have recently proposed a theoretical
explanation of the switching mechanism, actually observed in
different molecular junctions,$\cite{Blum05}$ based on the
electron-phonon interaction. Understanding and controlling the
effect of this interaction onto the electric current through
molecular devices is not only important for the field of molecular
electronics but establishes a strong link also to the physics of
Nano-ElectroMechanical Systems
(NEMS).$\cite{Blencowe05a,LaHaye04}$ Recent experiments show the
possibility to use single electron transistors coupled to a
mechanical oscillator as high sensitive
position$\cite{Blencowe05b,Knobel05}$ and mass$\cite{Knobel08}$
sensors. One of the challenges is to understand, control and use
the interplay between a quantum \`{}\`{}detector\'{}\'{} (electron
transistor) and a classical mechanical system.

The simplest molecular conduction junction comprises two metallic
electrodes connected by a single molecule. Such a junction,
including the effect of electron-phonon interaction, can be
described by the Anderson-Holstein (AH) model.$\cite{Hol}$ The
molecule is represented by one electronic level interacting
linearly with a local vibrational degree of freedom and connected
through tunneling with free-electron metals. Electron transport
within this model has received a lot of theoretical
attention.$\cite{MitraMillis,Galperin07}$ Despite the conceptual
simplicity, it gives rise to a very rich physics. Several
approaches have been adopted depending on the relative energy
scales in the problem. When the characteristic frequency of the
oscillator $\omega_{0}$ is of the same order of magnitude of the
tunneling frequency of the electrons on the molecule ($\sim
\Gamma$), a quantum treatment of oscillator dynamics is necessary.
In this case, it is useful to consider separately the limits of
weak and strong electron-phonon interaction strength relative to
the coupling of the level to the leads. The former corresponds to
nonresonant phonon-assisted electron tunneling, mostly encountered
in experiments in inelastic electron tunneling
spectroscopy$\cite{Wang}$ (IETS), and theoretically understood
within
%Landauer-Buttiker$\cite{Buttiker,Datta}$ and
Non Equilibrium Green Function (NEGF)
formalism.$\cite{Datta,Keld,Haug,Rynd}$ In the case of stronger
effective electron-phonon coupling, the perturbative treatment
breaks down, the conduction shows phononic blockade at small bias
(Franck-Condon effect$\cite{Koch05,Koch06,Cav}$) and one can
observe phonon sidebands in the conductance
spectra.$\cite{Flensberg}$

 The interpolation from weak to strong electron-phonon
coupling regime in a full quantum description of electron and
phonon subsystems is a very challenging problem. However, when the
vibrational motion is slow compared to the electronic tunneling
rate ($\omega_{0}<<\Gamma$), one can apply a simplified scheme in
the spirit of Born-Oppenheimer approximation. Indeed, in the
zero-order static theory (adiabatic ratio $\omega_0/\Gamma\mapsto
0$, oscillator mass $m \mapsto \infty$), one neglects the kinetic
energy of the oscillator and obtains an exact electronic problem
where the oscillator position enters as a parameter to be
determined self-consistently.$\cite{Galperin05}$ Some
authors$\cite{Mart,Pisto,Brandes}$ have already considered the
possibility to construct corrections to this picture, within the
AH model, in different parameter regimes and with different
techniques. Using the action functional formalism on the Keldysh
contour for the full interacting electron-phonon problem, Mozyrsky
\emph{et al.}$\cite{Mart}$ obtain an effective Langevin equation
for the oscillator field in the limit where the electron-leads are
considered as zero temperature thermostats. It comprises a
position-dependent dissipation term and white noise force. In the
strong electron-oscillator coupling regime, where the model shows
bistability, they find that the oscillator field acquires an
effective temperature linearly related to the bias, $V_{bias}$, as
a consequence of the coupling to the electronic bath. Pistolesi
\emph{et al.},$\cite{Pisto}$ starting from the Langevin equation,
generalize the previous results solving numerically the
Fokker-Plank equation associated with it, focusing again on the
extremely strong coupling regime. They obtain the dependence of
the current on the transport and gate voltages, as well as address
the problem of mechanical switching between the metastable states
of the oscillator potential.

A similar approach, based on the Feynman-Vernon influence
functional, was recently adopted by Hussein \emph{et
al.}$\cite{Brandes}$ obtaining the same Langevin equation.
However, they do not solve this equation and study the phase space
portraits of the Newton equation obtained in absence of electronic
bath induced noise. Moreover, they extend this analysis to a
molecular double dot molecular Hamiltonian. For both cases they
have explained the features of effective potential and friction
terms entering the oscillator equation of motion.

The approaches discussed above, based on an expansion of the
action in the adiabatic limit, are not able to disentangle the
origin of the quantum effects in the Langevin equation for the
oscillator. We propose here an alternative and more direct way
that allows us to clarify this point.

We construct systematically the dynamical finite mass corrections
to the static theory and their influence on the transport problem
in the following way: we perform an \emph{adiabatic expansion} on
the electron-oscillator self-energy following the approach of
Ref.$26$, obtaining a corresponding expansion of the Green
function. This expansion gives rise to the same friction and
fluctuating terms obtained with action functional techniques, but
clarifies that the quantum effects \`{}hidden\'{} in the
stochastic equation come only from the electronic subsystem.

We numerically solve the Langevin equation deriving the position
and velocity distribution probabilities of the oscillator for a
very large range of the relevant parameters. We find that, at
intermediate bias voltages, the velocity distributions $P(v)$
deviate from a gaussian shape as a result of the coupling of the
oscillator with the out-of-equilibrium electronic
bath.$\cite{Cav1}$ Correspondingly, the kinetic energy of the
oscillator shows a non monotonic behaviour as function of the bias
due to the slight change of the force exerted on the oscillator.

We study transport properties like the current-voltage
characteristic and the conductance, observing in the AH model a
dynamical reduction of the the \`{}polaronic\'{} shift and the
broadening of the electronic resonance due to the average over the
nonequilibrium position distribution probability of the
oscillator. We note an interesting strong enhancement of current
in the non gaussian region at intermediate bias, where the
infinite mass approximation prescribes very small conduction.

It is of paramount importance to study the range of validity of
the adiabatic approach. Making a thorough investigation of this
issue is crucial in order to get a match with experimental results
and exact theoretical calculations. We establish the range of
validity of the adiabatic approximation analyzing systematically
the behaviour of the oscillator's kinetic energy. We are able to
build up a diagram for the validity of classical approximation,
identifying Quantum (QR), Classical-Adiabatic (CAR) and Classical
Non-Adiabatic (CNAR) Regions. We compare the classical kinetic
energy of the oscillator with the Debye temperature
($k_{B}T_{D}\sim \hbar \omega_{0}$) to distinguish between QR
against CAR regimes, and with electron energy scale ($\sim
\hbar\Gamma$) to distinguish between CAR against CNAR regimes.

 We extend this analysis to the case of a
molecular Hamiltonian composed by a couple of sites interacting
with a single vibrational mode in the SSH
model.$\cite{SSH,SSH1,Asai,Ness,Kaat}$ In this case, because of
the direct coupling of the electron-oscillator interaction to the
intermolecular hopping, one expects that the role of the dynamical
fluctuations becomes crucial to determine the correct features of
the observables inherent to the transport problem.

In the limit of symmetric coupling with leads, we are able to
construct again a Langevin equation for the oscillator dynamics,
very similar to that derived in AH model. In this case, it is
possible to study the effect of the dynamics of the classical
vibrational mode on the electron hopping through the two molecular
sites. Vice-versa, one can also study the effect of intermolecular
electron degree of coherence onto the vibrational dynamics.

The new intermolecular electronic hopping scale $t$ introduces a
reduction of the CAR in the validity diagram. This reduction
becomes important if $t/\hbar\Gamma>1$ with the occurrence of QRs.
These new features are due to the stronger nonmonotonic behaviour
of the kinetic energy of the oscillator as function of the bias
voltage. For sufficiently large $t/\hbar\Gamma$, small interaction
strengths and at intermediate bias (non gaussian regime), the
kinetic energy curves show well defined minima. We point out a
correspondence between these minima and the maxima of the
conductance. Again, as concerns the electronic transport
properties, the dynamical fluctuations favour the conduction far
from the two electronic (static) resonances. In the SSH model, as
already stressed above, the effects of the dynamical fluctuations
become even more important. One can observe again a complete
erasing of the bistability and hysteretic behaviour predicted by
the infinite mass approximation.

 The paper is organized as follows: In Sec. II
we present the single level case within the AH
model. In Sec. III we deal with the case of two-site-SSH model.
%\begin{center}
%\textbf{II.The Anderson-Holstein model (AHM)}
%\end{center}
\section{The Anderson-Holstein (AH) model}
The spinless Anderson-Holstein model is the simplest model of a
molecular junction including the effect of electron-phonon
interaction. The molecule is modeled as a single electronic level
interacting locally with a single vibrational mode. The electronic
system is described by the standard junction Hamiltonian
$\hat{\cal H}_{el}={\hat H}_{mol}+ {\hat H}_{tun} +{\hat
H}_{leads}$, with
\begin{equation}
{\hat H}_{mol}=E_{g}{\hat d^{\dag}}{\hat d},
\end{equation}
\begin{equation}
{\hat H}_{tun}=\sum_{k,\alpha}(V_{k,\alpha}{\hat
c^{\dag}_{k,\alpha}}{\hat d}+ h.c.),\label{HTun}
\end{equation}
\begin{equation}
{\hat H}_{leads}=\sum_{k,\alpha}\varepsilon_{k,\alpha}{\hat
c^{\dag}_{k,\alpha}}{\hat c_{k,\alpha}}.\label{Hleads}
\end{equation}

The molecular electronic level has energy $E_{g}$ and creation
(annihilation) operators ${\hat d^{\dag}} ({\hat d})$. The
operators ${\hat c^{\dag}_{k,\alpha}} ({\hat c}_{k,\alpha})$
create (annihilate) electrons with momentum $k$ and energy
$\varepsilon_{k,\alpha}=\xi_{k,\alpha}-\mu_{\alpha}$ in the left
($\alpha=L$) or right ($\alpha=R$) free metallic leads. The
chemical potentials in the leads, $\mu_{L}$ and $\mu_{R}$, are
assumed to be biased by an external voltage,
$eV_{bias}=\mu_{L}-\mu_{R}$. Electronic tunneling between the
molecular dot and a state in the lead has amplitude
$V_{k,\alpha}$. We consider the oscillator dynamics
\`{}classical\'{} from the beginning and described by the position
and momentum variables $x, p$.

The Hamiltonian of the oscillator is given by
\begin{equation}
H_{osc}={p^{2}\over 2m} + {1\over 2}m
\omega^{2}_{0}x^{2},\label{Hosc}
\end{equation}
characterized by the frequency $\omega_{0}$ and the effective mass
$m$. The interaction (typically of electrostatic origin) is
provided by a simple linear coupling between the electron
occupation on the molecule, ${\hat d^{\dag}}{\hat d}$, and the
displacement of the oscillator
\begin{equation}
{\hat H}_{int}=\lambda x{\hat d^{\dag}}{\hat d},\label{Hint}
\end{equation}
where $\lambda$ is the electron-oscillator coupling (EOC)
strength. The overall Hamiltonian is therefore given by $\hat{\cal
H}=\hat{\cal H}_{el}+ H_{osc} +{\hat H}_{int}$.

In the following, the coupling between the electron system and the
vibrational mode will be often described in terms of the
electron-phonon coupling energy
$E_{p}=\lambda^2/(2m\omega_{0}^{2})$, while the coupling to the
leads by the tunneling rate
$\Gamma_{k,\alpha}=2\pi\rho_{\alpha}|V_{k,\alpha}|^{2}/\hbar$ (the
full hybridization width of the molecular orbital is then
$\hbar\Gamma_{k}=\hbar\Gamma_{k,L} +\hbar\Gamma_{k,R}$), where
$\rho_{\alpha}$ is the density of states of the lead $\alpha$. For
the sake of simplicity, we will suppose flat density of states for
the leads within the wide-band approximation
$(\Gamma_{k,\alpha}\mapsto \Gamma_{\alpha})$. In this paper we
will measure length in units of $x_{0}={\lambda\over
m\omega_{0}^{2}}$ and energy in units of $\hbar\Gamma$. Finally,
the leads will be considered as zero temperature thermostats.

In the next subsections, we will first (subsection A) analyze the
coupled electron-oscillator problem in the limit of infinite mass
for the oscillator. We will then indicate how to construct
(subsection B) the stochastic Langevin equation for the dynamics
of the oscillator including the finite mass effect. In the
subsection C we will solve numerically the stochastic equation and
analyze the effects of the oscillator dynamics on the electronic
observables inherent to the transport problem (I-V characteristic
and conductance).
%\begin{center}
%\textbf{II.A Out of equilibrium Born-Opphenheimer approximation:
%generalized static potential}
%\end{center}

\subsection{Out of equilibrium Born-Oppenheimer approximation:
infinite mass (static) case} When the vibrational motion of the
oscillator is slow with respect to all electronic energy scales,
it is possible to decouple oscillator and electronic dynamics. In
the spirit of Born-Oppenheimer approximation, we consider the
limit $m\mapsto \infty$ in the full Hamiltonian disregarding the
kinetic energy of the oscillator. The electronic dynamics is
therefore equivalent to a non-interacting resonant single level
problem with energy level renormalized by the \`{}polaronic\'{}
shift $E_{g}\mapsto E_{g}+\lambda x$. The retarded (advanced)
Green functions $G^{r(a)}(\omega,x)$ and the lesser (greater)
Green functions $G^{<(>)}(\omega,x)$ in stationary nonequilibrium
conditions are derived within the Keldysh
formalism$\cite{Keld,Haug}$ and depend parametrically by the
displacement coordinate $x$. Starting from the force exerted on
the oscillator
\begin{equation}\label{for}
F=-m\omega_{0}^{2}x+\lambda\langle
\hat n_{el}\rangle(x),
\end{equation}
where
\begin{equation}\label{occu}
\langle \hat
n_{el}\rangle(x)=\sum_{\alpha=L,R}\hbar^{2}\Gamma_{\alpha}\int
{d\omega\over2\pi} f_{\alpha}(\omega)|G^{r}(\omega,x)|^{2},
\end{equation}
with $f_{\alpha}(\omega)$ Fermi function of the lead $\alpha=R,L$,
one can therefore straightforwardly compute the expression of the
generalized potential in nonequilibrium conditions (obtained
applying symmetric bias unbalance $\mu_{R}=-eV_{bias}/2$,
$\mu_{L}=eV_{bias}/2$)
\begin{eqnarray}\label{enpotNE1sito}
&&U(x)={1\over 2}m\omega_{0}^{2}x^{2} +{\lambda x\over 2}-
\sum_{\alpha=L,R}\Bigg[{\mu_{\alpha}-E_{g}-\lambda x\over
2\pi}\times \nonumber\\
&&\arctan\Big({\mu_{\alpha}-E_{g}-\lambda x\over
\hbar\Gamma/2}\Big) - {\hbar\Gamma\over 8\pi}
\ln[4(\mu_{\alpha}-E_{g}-\lambda
x)^2\nonumber\\
&&+(\hbar\Gamma)^2]\Bigg].
\end{eqnarray}

\begin{figure}
\centering
{\includegraphics[width=8cm,height=9.0cm,angle=-90]{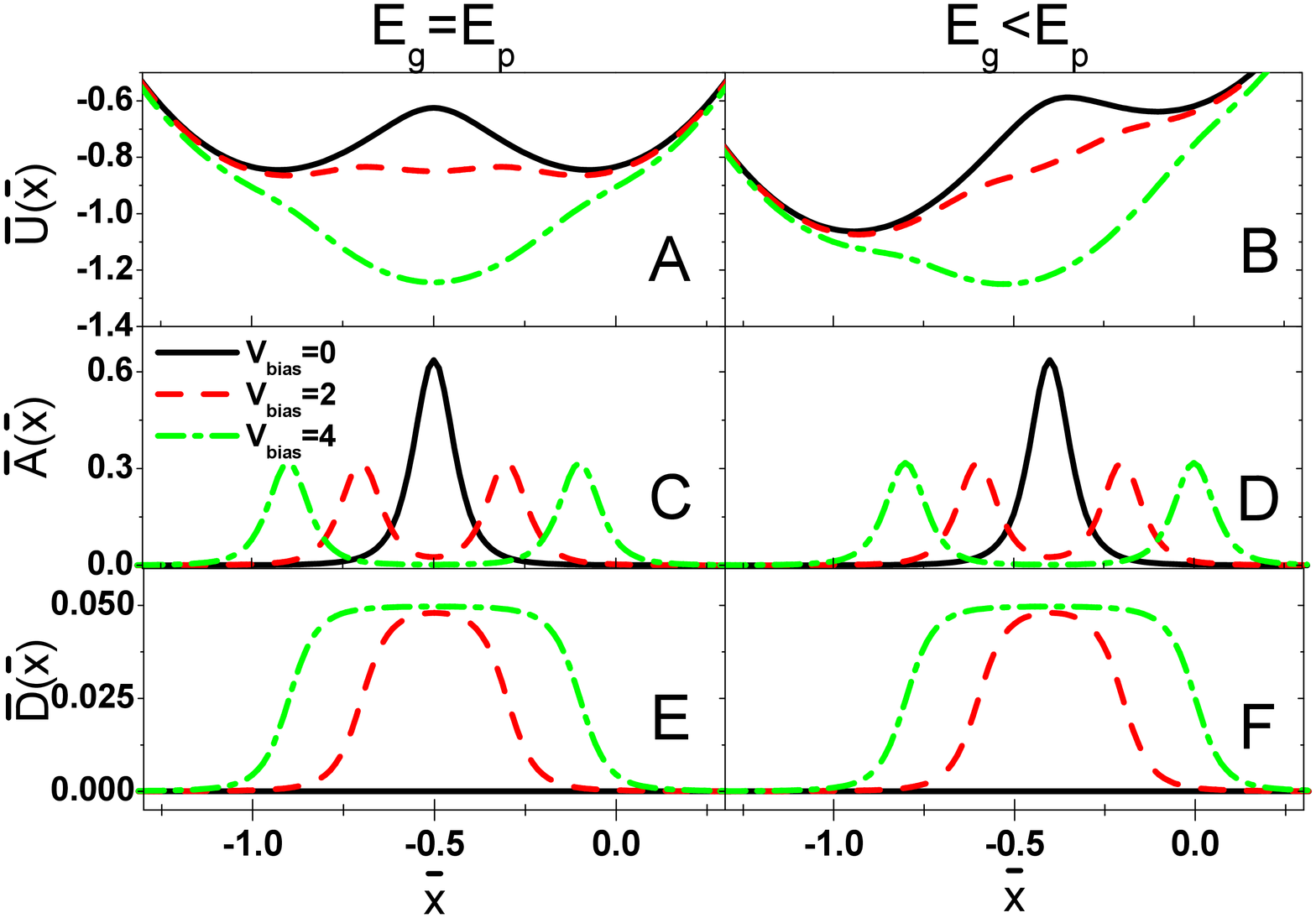}}
\caption{(Color online). Spatial dependence of the dimensionless
generalized static potential ${\bar U}({\bar x})$ (panels A,B),
friction coefficient ${\bar A}({\bar x})$ (panels C,D),
fluctuating term ${\bar D}({\bar x})$ (panels E,F) for symmetric
$E_{g}\sim E_{p}$ and asymmetric $E_{g}<E_{p}$ minima and
different values of bias, $V_{bias}=0$ (solid (black) curve),
$V_{bias}=2$ (dashed (red) curve), $V_{bias}=4$ (dashed dot
(green) curve). The potential is expressed in $\hbar\Gamma$ units
(${\bar U}=U/\hbar\Gamma$), the friction coefficient in
$m\omega_{0}$ units (${\bar A}=A/m\omega_{0}$), the fluctuating
term in $\lambda^{2}/\omega_{0}$ units, (${\bar
D}=D/(\lambda^{2}/\omega_{0})$). $V_{bias}$ values are expressed
in $\hbar\Gamma/e$ units, where $e$ is the electron charge. The
dimensionless position variable ${\bar x}$ is defined as ${\bar
x}=x/x_{0}$ with $x_{0}={\lambda\over
m\omega_{0}^{2}}$.}\label{potstat1sito}
\end{figure}
This generalized oscillator potential depends parametrically by
the spring constant $m\omega_{0}^{2}=k$, the EOC strength
$\lambda$, the energy of the electron level $E_{g}$ (which can be
considered a gate potential), the coupling to the leads $\Gamma$
and finally by the bias, $V_{bias}$. In Fig.$\ref{potstat1sito}$
(panels A,B), we present some features of the generalized
potential $U(x)$ in the strong coupling regime
($E_{p}>\hbar\Gamma$), where the potential shows several minima.
For $E_{g}\sim E_{p}$ and not too large bias (panel A,
Fig.$\ref{potstat1sito}$), the potential develops two symmetric
minima near $x\simeq 0$ (corresponding to $\langle {\hat
n}_{el}\rangle\simeq 0$) and $x\simeq -1$ (corresponding to
$\langle {\hat n}_{el}\rangle\simeq 1$) separated by a barrier
whose height is roughly proportional to $E_{p}$. This bistable
regime corresponds to the physical situation where the bare
electron level $E_{g}$ is above the chemical potential of both
leads, while the renormalized charged level $E_{g}-2E_{p}$ is
below them. The molecule can stay in one minimum or in the other.
If we increase the bias $V_{bias}$, the potential $U(x)$ shows a
third minimum corresponding to average electron occupation on the
molecule $\langle {\hat n}_{el}\rangle\simeq 1/2$ and, for
sufficiently large $V_{bias}$, only this minimum remains. If
$E_{g}< E_{p}$ the potential also shows two or more minima but
they are asymmetric (panel B, Fig.$\ref{potstat1sito}$). For
sufficiently large bias, the common feature is the existence of a
single minimum corresponding to occupation $\langle {\hat
n}_{el}\rangle\simeq 1/2$.

In the above analysis the displacement $x$ has been used as a free
parameter. Actually, the only $x$ values relevant for the
electronic properties in the static approximation are those which
solve the equation $F(x)=0$. These solutions depend parametrically
by all the parameters of the theory (in particular by the bias
$V_{bias}$). This may yield transitions between different local
minima in the potential, determining in the electronic
current-voltage characteristic the onset of interesting non linear
phenomena like hysteresis, bistability and Negative Differential
Resistance (NDR).$\cite{Galperin05}$ Indeed, the authors of
Ref.$4$ proposed a polaron mechanism within the AH model to
explain such phenomena, effectively observed in transport
experiments on molecular devices. However, the results of the
static approximation can be strongly modified by dynamical
effects. Indeed, corrections due to the finite (though large) mass
of the oscillator are expected to be
important.$\cite{Mart,Pisto,Brandes}$ As we shall see in the next
sections, the inclusion of the finite mass effect on the
oscillator dynamics gives rise to a stochastic Langevin equation
with a position dependent dissipation term and white noise force.
The stochastic fluctuations of the oscillator motion will strongly
modify the current-voltage characteristics obtained in the
infinite mass approximation.

\subsection{Dynamical (finite mass) corrections to static case: setting
Langevin equation for the oscillator}

Within the static approximation (infinite mass), the main effect
of the nonequilibrium fast electronic environment is the
modification of the force (Eq.$(\ref{for})$) experienced by the
mechanical oscillator. The oscillator has no dynamics at all and
the displacements $x$ are \`{}frozen\'{} in suitable points of the
configuration space given by the equation $F(x)=0$. In this
section, we show how to include dynamical corrections due to
finite mass of the oscillator.

First of all, we should include the time dependence of the
oscillator dynamics in the Hamiltonian of the electronic problem
(where we still neglect the oscillator kinetic energy, $\hat{\cal
H} \mapsto \hat{\cal H}(x(t))$). Using the extension of the
Keldysh formalism to time dependent cases,$\cite{Haug}$ we can
solve the Dyson and Keldysh equations for the molecular Green
functions which now depends on times $t$ and $t'$ separately. The
retarded molecular Green function can be obtained analitically
(with $E_{g}(t)=E_{g}+\lambda x(t)$)
\begin{equation}
G^{r}(t,t')=-{\imath\over\hbar} \theta(t-t')
e^{-\imath\int_{t'}^{t}dt_{1}({E_{g}(t_{1})\over
\hbar}-\imath\Gamma/2)},\label{grtimedep}
\end{equation}
and depends in non linear way by the entire dynamics $x(t)$ of the
oscillator. Indeed, the dependence by the entire history of the
oscillator motion prevents us from Fourier transforming the Green
function and leads to an intractable problem.

In order to overcome this difficulty, we resort to an adiabatic
approximation of the molecular Green function. As thoroughly
discussed in the Appendix $A$, one should restart from the Dyson
equation for the Green function (Eq.$(\ref{DysonT})$) and perform
an expansion in the electron-phonon self-energy where a separation
between \`{}slow\'{} and \`{}fast\'{} time scales was preliminary
accomplished (Eqs.$(\ref{self}$,$\ref{selfexpansion})$). The
expansion, performed with respect to the \`{}slow\'{} time
$(t+t')/2$, allows to disentangle the non-local time dependence of
the Green function. In the end, the truncated Green function can
be Fourier transformed with respect to \`{}fast\'{} time $t-t'$,
gaining a \`{}slow\'{} time dependence and a linear correction in
the oscillator velocity (See
Eqs.$(\ref{AdiExpansion1}$,$\ref{G0adi}$,$\ref{G1adi})$).

\subsubsection{Abiabatic Approximation: calculation of damping and fluctuating term}
We have now the tools to calculate the adiabatic corrections to
the force acting on the mechanical oscillator. For the sake of
clarity, we rewrite here the adiabatic expansion of the Fourier
transformed molecular Green function derived in the Appendix A
(Eq.$\ref{AdiExpansion1}$)
\begin{equation}
G^{r}(\omega,t)\simeq G^{r}_{0}(\omega,t)+G^{r}_{1}(\omega,t).
\end{equation}
The explicit expressions of $G^{r}_{0}$ and $G^{r}_{1}$ are given
by Eqs.$(\ref{G0adi}$,$\ref{G1adi})$. Actually, in order to
introduce dynamical effects on the force Eq.$(\ref{for})$, we have
to calculate the adiabatic corrections to the lesser-Green
function that is directly related to the occupation
\begin{eqnarray}
&&\langle \hat n\rangle(t)= -\imath \hbar G^{<}(t,t)=\nonumber\\
&&\sum_{\alpha=L,R}\hbar^{2}\Gamma_{\alpha}\int {d\omega\over2\pi}
f_{\alpha}(\omega)|G^{r}(\omega,t)|^{2}.\label{glestimedep}
\end{eqnarray}
%where $G^{r}(\omega, t)$ is given by
%\begin{equation}
%G^{r}(\omega, t)=\int d\tau G^{r}(t,\tau)e^{\imath \omega
%\tau}.\label{Btimedep}
%\end{equation}
From Eq.$(\ref{G0adi})$, one obtains, at zero-order, an expression
for the occupation of the same form of the static limit
(Eq.(\ref{occu})) with the substitution $E_{g} \leftrightarrow
E_{g}(t)$, acquiring a weak time dependence through the slow
variable $t$. Adding the first order correction Eq.$(\ref{G1adi})$
into the Eq.$(\ref{glestimedep})$, and neglecting terms
proportional to the square velocity of the oscillator, we obtain
\begin{eqnarray}
&&\langle \hat n\rangle (t)\simeq
\sum_{\alpha=L,R}\hbar^{2}\Gamma_{\alpha}\int {d\omega\over 2\pi}
f_{\alpha}(\omega)\Bigg(1+\nonumber\\
&&{\hbar\Gamma\over 2}{\partial E_{g}\over \partial t}
|G^{r}_{0}(\omega,t)|^{2}{\partial \over
\partial \omega}\Bigg)|G^{r}_{0}(\omega,t)|^{2}.\label{occuadiabatico}
\end{eqnarray}
In the end, the force given by Eq.($\ref{for}$) modifies to
\begin{equation}\label{forMOD}
F(x) \mapsto F'(x,v)=F(x)- A(x)v,
\end{equation}
where $v=\dot{x}$ is the velocity of the oscillator. We can
conclude that, both in equilibrium and in out-of-equilibrium
conditions, the interaction with the leads introduces a
dissipative correction term to the oscillator dynamics.

Now, we observe that the introduction of a dissipative term cannot
be the unique dynamical effect for a classical dynamical system in
contact with an environment. It is well known that a fluctuating
term should be included to take correctly into account the effect
of the bath degree's of freedom. In our case, in order to include
\emph{completely} the effect of the \`{}fast\'{} electronic
environment on the oscillator motion, we propose to take into
account the fluctuations of the force$\cite{Messina}$ acting on
the oscillator. These are induced by the intrisic \`{}quantum\'{}
fluctuations of the electronic subsystem.

We add to the \emph{average} force contribution Eq.$(\ref{for})$,
suitable corrected by the damping term, Eq.$(\ref{forMOD})$, a
stochastic fluctuating term able to take into account the effect
of the electronic quantum fluctuations on the classical dynamics
of the oscillator. Indeed, because we are considering
zero-temperature leads, one expects that these fluctuations are
triggered by a finite bias voltage applied to the junction. As we
shall see in the next sections, our approach is very accurate for
junctions driven by a finite bias voltages, but it is far less
accurate to describe the physics in the small bias (quantum)
regime.

We estimate the noise strength evaluating the \emph{average} of
the square fluctuation of the force over the electronic steady
state. This fluctuating term is directly related to the
fluctuation of the electron occupation
\begin{eqnarray}\label{fluforce}
&&\langle\delta\hat F(t)\delta\hat F(t')\rangle=
\lambda^{2}\langle\delta\hat n(t)\delta\hat n(t')\rangle
= \nonumber\\
&&\lambda^{2}\Big(\langle\hat n(t)\hat n(t')\rangle-\langle\hat
n(t)\rangle\langle\hat n(t')\rangle\Big).
\end{eqnarray}
Decoupling the term $\langle\hat n(t)\hat n(t')\rangle$ with the
Wick theorem, one obtains
\begin{eqnarray}\label{fluforce}
&&\langle\delta\hat F(t)\delta\hat
F(t')\rangle=\lambda^{2}\hbar^{2}G^{<}(t'-t)G^{>}(t-t')=\nonumber\\
&=&\lambda^{2}\hbar^{2}G_{0}^{<}(t'-t)G_{0}^{>}(t-t'),
\end{eqnarray}
where we have used zero-order time-dependent Green functions (as
in Eq. $(\ref{G0adi})$) in order to take only first order
corrections in the adiabatic ratio ${\omega_{0}\over \Gamma}$. At
this level of approximation, we have obtained a multiplicative
coloured noise.$\cite{lange}$ According to the adiabatic
approximation, we can further simplify the fluctuating term
retaining only the zero-frequency component of the noise
\begin{eqnarray}\label{noise1}
&&\lim_{\varepsilon\mapsto 0}\lambda^{2}\hbar^{2}\int d\varepsilon
e^{\imath \varepsilon (t-t')}\int {d\omega\over 2\pi}
G^{<}_{0}(\omega+\varepsilon)G^{>}_{0}(\omega)\nonumber\\
&&\simeq D(x) \delta(t-t'),
\end{eqnarray}
corresponding to electronic times scales comparable with that of
the oscillator. In this way, one obtains a multiplicative white
noise term in the equation of motion for the
oscillator.$\cite{lange}$ The resulting Langevin equation for the
oscillator dynamics becomes
\begin{eqnarray}\label{Langevin1}
m \ddot{x} &+& A(x)\dot{x}=F(x)+ \sqrt{D(x)}\xi(t),\\
\langle\xi(t)\rangle&=&0,\;\;\;\;\langle\xi(t)\xi(t')\rangle=\delta(t-t'),\nonumber
\end{eqnarray}
where $\xi(t)$ is a standard white noise term. Explicitly, the
damping term $A(x)$ is given by (from Eq.$(\ref{occuadiabatico})$)
\begin{equation}\label{damping}
A(x)={4m\omega_{0}\over\pi}{\hbar\omega_{0}\over\hbar\Gamma}{E_{p}\over\hbar\Gamma}
\sum_{\alpha=L,-R}\Bigg({1\over [({\mu_{\alpha}-E_{g}-\lambda
x\over \hbar\Gamma})^{2}+1]^{2}}\Bigg),
\end{equation}
while the fluctuating term is (from Eq.$(\ref{noise1})$)
\begin{eqnarray}\label{noise}
D(x)&=&{m\omega_{0} E_{p}\over
\pi}{\hbar\omega_{0}\over\hbar\Gamma}
\sum_{\alpha=L,-R}\Bigg(\arctan({\mu_{\alpha}-E_{g}-\lambda
x\over \hbar\Gamma})\nonumber\\
&+&{{\mu_{\alpha}-E_{g}-\lambda x\over \hbar\Gamma}\over
[({\mu_{\alpha}-E_{g}-\lambda x\over \hbar\Gamma})^{2}+1]}\Bigg),
\end{eqnarray}
where it is understood that $\sum_{\alpha=L,-R}
K(\alpha)=K(L)-K(R)$, for a generic function $K(\alpha)$. We note
that Eqs.$(\ref{Langevin1})$, $(\ref{damping})$ and
$(\ref{noise})$ are identical to that obtained in Ref.$23-25$.
Introducing a natural temporal unit $t_{0}=1/\omega_{0}$, the
dimensionless damping $\bar A (\bar x)$ and fluctuating $\bar D
(\bar x)$ coefficients result proportional to the adiabatic ratio
$\omega_{0}/\Gamma$. As concerns the spatial dependence of the
damping term, one can note in Fig.$\ref{potstat1sito}$ (panels
C-D) that it is almost localized on the position of the local
minima of the static potential. The fluctuating coefficient, as
shown in Fig.$\ref{potstat1sito}$ (solid (black) line in panels
E-F), vanishes at equilibrium (bias voltage $V_{bias}=0$). Only
applying finite bias it becomes different from zero. In
Fig.$\ref{potstat1sito}$ (dashed (red) and dashed dotted (green)
lines in panels E-F), we show that its spatial extension increases
as the bias increases. Finally, one can note that $A(x)$ and
$D(x)$ are almost independent by the ratio $E_{g}/E_{p}$.

\subsection{Numerical solution of Langevin equation:
\emph{electronic} observables and limits of the
stochastic approach}

From the Langevin equation Eq.$(\ref{Langevin1})$ it is possible
to derive the distribution probabilities $P(x)$ and $P(v)$ for the
position and velocities variables of the oscillator. We have
evaluated them solving the second order stochastic differential
equation with the 4-order stochastic Runge-Kutta algorithm
developed by R. L. Honeycutt.$\cite{HoneyI,HoneyII}$ First of all,
as suggested in Ref.$37$, in order to solve our second order
ordinary differential equation with multiplicative white noise, we
decompose the problem in a set of three first order differential
equations. The third one takes into account the effect of spatial
dependence of the noise involving a non multiplicative noise term.
For our simulations we have fixed a time step $t_{s}=0.1\tau$
($\tau=1/\omega_{0}$) and set long simulation times up to
$T=10^{9}t_{s}$. Within these settings, the algorithm shows an
excellent stability in the whole range of model parameters. In
order to construct our histograms, we have sampled the values of
$x(t)$ and $v(t)$ every $100$ time steps. We have therefore
obtained the distribution probabilities for the stationary state
of the oscillator.

Given our assumption about the separation between the slow ionic
(vibrational) and fast electronic (tunneling) timescales, the
problem of evaluating a generic observable (electronic or not) of
the system reduces to the evaluation of that quantity for a fixed
position $x$ and velocity $v$ of the oscillator, with the
consequent averaging over the stationary probability
distributions, $P(x)$ and $P(v)$. Therefore, for a generic
observable which depends only on position, $O(x)$, the averaged
quantity is
\begin{eqnarray}
\langle O(x)\rangle &=& \int dx P(x) O(x),
\end{eqnarray}
while, for an observable which depends on velocity variable only,
one has
\begin{eqnarray}
\langle O(v)\rangle &=& \int dv P(v) O(v).
\end{eqnarray}
In our case, the current, the
spectral function and the electronic occupation depend only on the
position variable
\begin{eqnarray}
I(x) &=& {e\over \hbar} \int_{\mu_{R}}^{\mu_{L}}{d\hbar\omega\over 2\pi}{\hbar\Gamma_{L}\hbar\Gamma_{R}\over \hbar\Gamma}A_{Spec}(\omega,x), \\
A_{Spec}(\omega,x)&=&{\hbar\Gamma\over(\hbar\omega-E_{g}-\lambda x)^{2} +\hbar^{2}\Gamma^{2}/4}, \\
\langle \hat n\rangle(x) &=& {1\over
2}+{1\over2\pi}\sum_{\alpha=R,L}\arctan\Bigg[{\mu_{\alpha}-E_{g}-\lambda
x\over\hbar\Gamma/2}\Bigg].\nonumber\\
\end{eqnarray}
\begin{figure}
\centering
{\includegraphics[width=8cm,height=9.0cm,angle=-90]{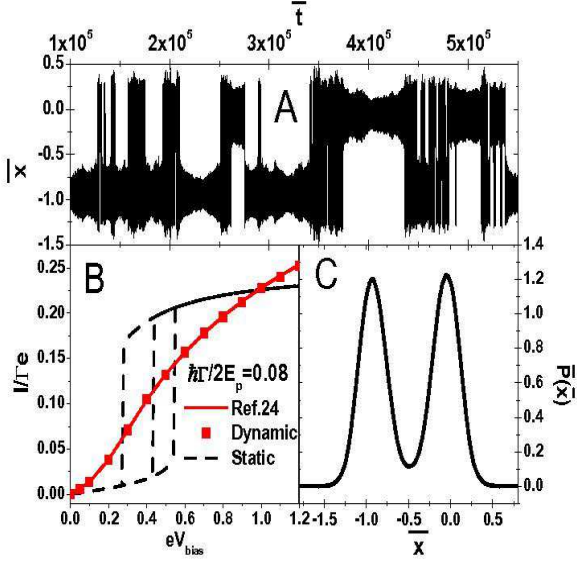}}
\caption{(Color online). Panel A: Solution of the Langevin
equation Eq.$(\ref{Langevin1})$ in the extremely strong coupling
regime $E_{p}>>\hbar\Gamma>>\hbar\omega_{0}$
($\hbar\omega_{0}/2E_{p}=10^{-3}$) for $\hbar\Gamma/2E_{p}=0.08$,
$E_{g}=E_{p}$ and $eV_{Bias}/2E_{p}=0.1$. Panel B: Current
($e\Gamma$ units) voltage ($eV_{bias}$ in $2 E_{p}$ units)
characteristic for the same value of $\hbar\Gamma/2E_{p}$ as
above. Solid (red) curve is drown from Ref.$24$, square line
indicates our dynamic simulation and dashed line indicates static
I-V. Panel C: Dimensionless position distribution probability for
the same values of parameters as in Panel A. The dimensionless
position ${\bar x}$, time ${\bar t}$ and distribution function
${\bar P}$ are defined as ${\bar x}=x/x_{0}$, ${\bar t}=t/t_{0}$,
${\bar P}=P/(1/x_{0})$, with $x_{0}={\lambda\over
m\omega_{0}^{2}}$ and $t_{0}=1/\omega_{0}$,
respectively.}\label{pisto}
\end{figure}
The position distribution probabilities $P(x)$ have been already
discussed by authors of Ref.$24$ in the extremely strong coupling
regime $E_{p}>>\hbar\Gamma>>\hbar\omega_{0}$. They analyze the
case where the static potential shows two symmetric or asymmetric
wells separated by a very high barrier. In this regime, solving
numerically the Fokker-Plank equation of the problem, they
estimate the switching-rates by evaluating the escape times from
each well of the generalized potential. Indeed, this point is
interesting for clarifying the role of electron-phonon interaction
in the appearance of a bistable behaviour in single molecule
tunneling devices. One of the results of this paper is that the
multistability and hysteretic behaviour in the current-voltage
characteristic disappear if the dynamical effects of the
oscillator motion are taken into account. To clarify this point,
we focus here on the case (already considered in Ref.$24$) where
the switching times between different oscillator potential wells
are very long, and the oscillator jumps between two states (see
panel A of Fig.$\ref{pisto}$) corresponding to very small
electronic currents. In order to explore the same regime of
parameters, in our approach very long simulation times as
$T=10^{9}t_{s}$ are necessary for sampling the entire phase space
experienced during the dynamics. Nevertheless, as shown in Panel B
of Fig.$\ref{pisto}$, we get an excellent agreement with Pistolesi
\emph{et al}. results. It is interesting to note that in the paper
of Ref.$24$, the authors consider a broadening $\Gamma$ which is
twice our values (we show in the caption of Fig.$\ref{pisto}$ the
comparison between the simulations taking correctly into account
this factor). In the small bias regime, we observe a strong
suppression of the current. The oscillator spends a long time in
each potential well, suddenly jumps into the other and then comes
back in the same way (see panel A in Fig.$\ref{pisto}$). For
clarity, we show in panel C of Fig.$\ref{pisto}$ the corresponding
position distribution probability $P(x)$. The maxima of $P(x)$
correspond to two small current carrying states: the position of
the molecular energy level is far above ($E_{g}\sim E_{p}$) or
below ($E_{g}\sim-E_{p}$) the chemical potential of the leads. For
sufficiently large bias voltage, as discussed in subsection A,
appears a third minimum in the static potential. This minimum
corresponds to a large-current carrying state determining a
continuous enhancement of the current, against the abrupt
discontinuity (or hysteresis) which would been obtained in the
static approximation (dashed line in panel B of
Fig.$\ref{pisto}$).
\begin{figure}
\centering
{\includegraphics[width=8cm,height=9.0cm,angle=-90]{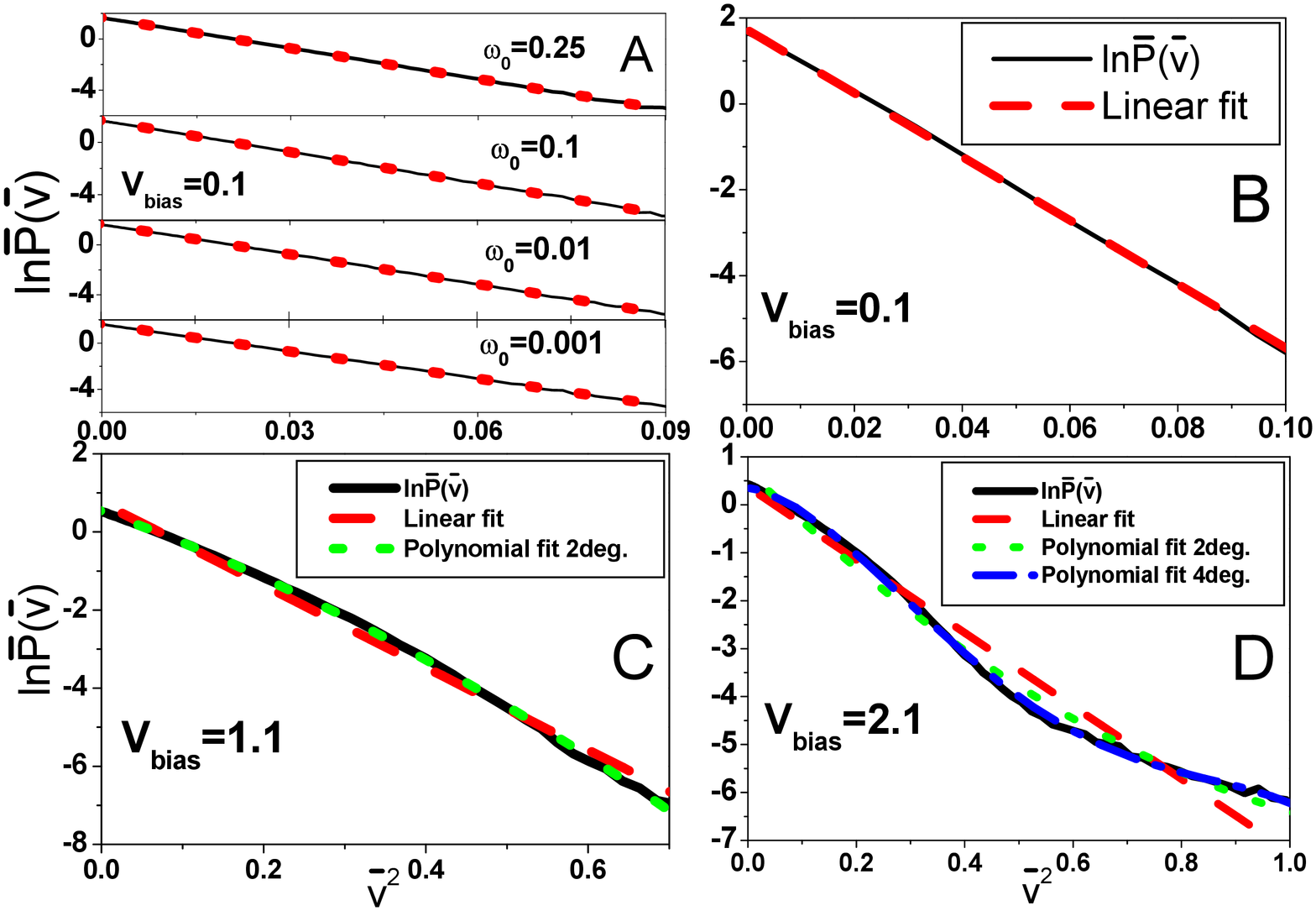}}
\caption{(Color online). Panel $A$: log-plot of
\emph{dimensionless} velocity probability distribution function
vs. $v^{2}$, at different adiabatic ratios (the values of
$\omega_{0}$ shown in the figure are in $\Gamma$ units), fixed
bias voltage $V_{bias}=0.1$ and different gate voltages and EOC
strengths (not shown in the graph). The dotted (red) lines
indicate that curves have a good linear fit. Panels $B-C-D$:
log-plot of velocity probability distribution function vs.$v^{2}$
for $V_{bias}=0.1$, $V_{bias}=1.1$, $V_{bias}=2.1$, respectively.
The dashed (red) line indicates linear fitting. Dotted (green) and
dash dotted (blue) lines indicate polynomial fitting of 2nd and
4th degree. $V_{bias}$ values are expressed in $\hbar\Gamma/e$
units. The dimensionless distribution function is defined as
${\bar P}=P/(m\omega_{0}/\lambda)$, while $v^{2}$ is expressed in
$(\lambda/m\omega_{0})^{2}$ units.}\label{logvqplot}
\end{figure}
\subsubsection{Non gaussian features of P(v) and study of the average kinetic energy of the oscillator}
\begin{figure}
\centering
{\includegraphics[width=8cm,height=9.0cm,angle=-90]{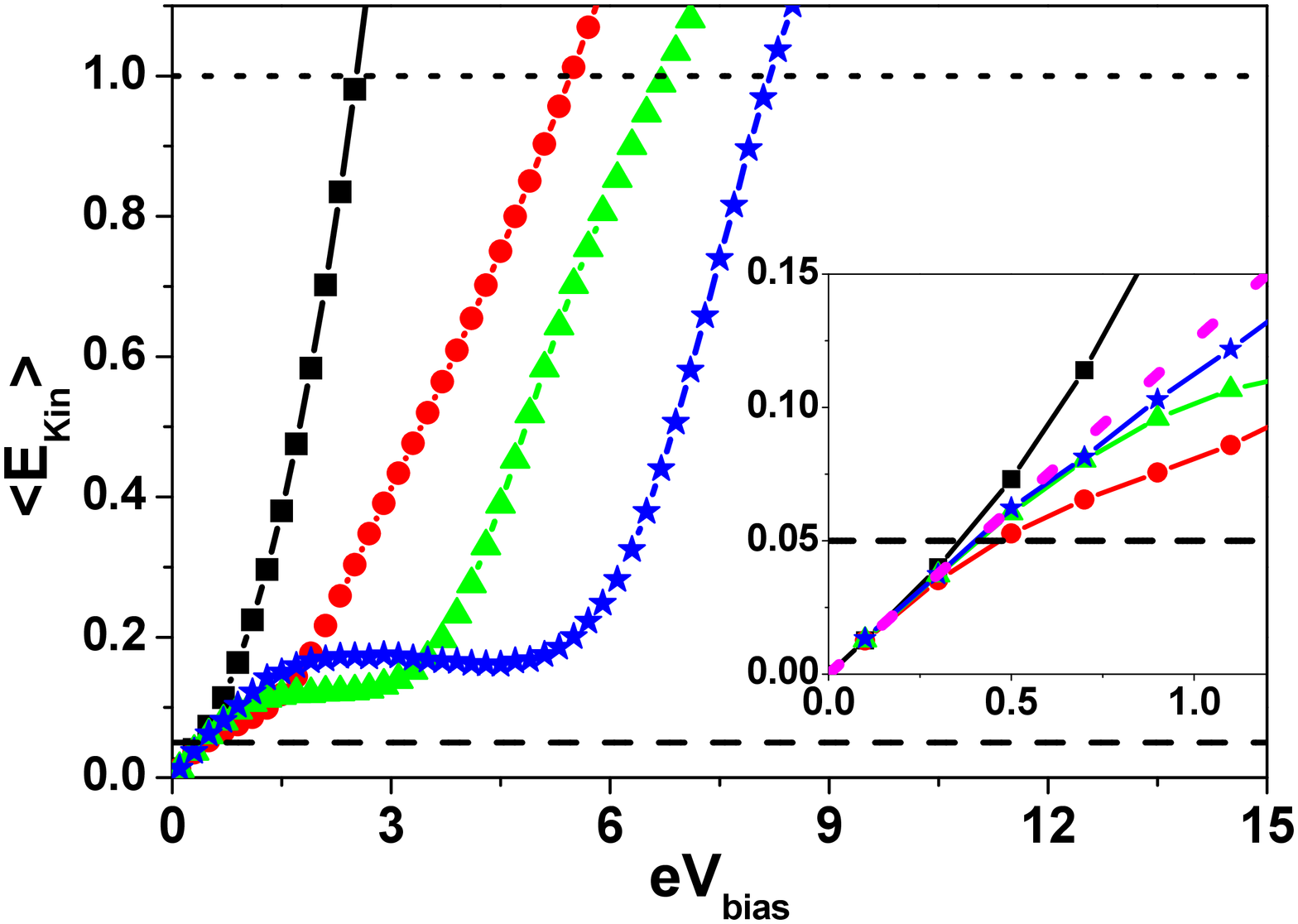}}
\caption{(Color online). Main: plot of average kinetic energy
$\langle E_{Kin}\rangle$ as function of the bias voltage at fixed
adiabatic ratio $\omega_{0}/\Gamma=0.1$ and gate voltage
$E_{g}=0$, for different interaction strengths $E_{p}$:
$E_{p}=0.1$ square (black) curve, $E_{p}=1.0$ circle (red) curve,
$E_{p}=2.0$ triangle (green) curve, $E_{p}=3.0$ star (blue) curve.
Two constant energy lines $E=\hbar \omega_{0}/2\hbar\Gamma=0.05$
(dashed) and $E=\hbar \Gamma=1$ (dotted) are also plotted. Inset:
Average kinetic energy $\langle E_{Kin}\rangle$ for small bias
voltages for the same parameter values of the main plot. The
dotted (magenta) line indicates the linear approximation
$eV_{bias}/8$ derived in Ref.$23$ (we choose a broadening
$\hbar\Gamma$ half that used in Ref.$23$). All the quantities
($\langle E_{Kin}\rangle$, $E_{p}$, $E_{g}$ and $eV_{bias}$) are
in unit $\hbar \Gamma$.}\label{Ecin111}
\end{figure}

In this section we focus our attention on the oscillator
observables $O(v)$ which depend on the velocity $v$. We remark
that the oscillator is coupled to the electronic bath only through
the interaction term ${\hat H}_{int}$, Eq.$(\ref{Hint})$. As the
bias voltage increases, this bath is strongly driven out of
equilibrium. It is therefore important to analyze the effect of
the electronic subsystem on the oscillator distribution
probability $P(v)$ as a function of the bias voltage. In the small
bias regime, regardless the value of the gate voltages $E_{g}$ and
the coupling $E_{p}$, as shown in Fig.$\ref{logvqplot}$ (panel
$A$) for different adiabatic ratios (from $\omega_{0}=10^{-3}$ to
$\omega_{0}=0.25$), the velocity distribution probabilities $P(v)$
are gaussian. In this regime, the nonequilibrium electronic bath
behaves like a conventional bath for the oscillator with an
 \`{}effective\'{} temperature linearly proportional to the bias voltage.
As described in the inset of Fig.$\ref{Ecin111}$, at arbitrary
$E_{p}$ and gate voltages the kinetic energy curves show a common
linear trend at small bias with a slope $V_{bias}/4$ in agreement
with Mozyrsky \emph{et al.} (we get $V_{bias}/8$ because we choose
a broadening $\hbar\Gamma$ half that used in Ref.$23$). As we
increase the bias voltage, the $(\log P(v)$ vs. $ v^{2})$ plot
starts to deviate from a linear trend, as shown in
Fig.$\ref{logvqplot}$, panels $B-C-D$. This behaviour indicates
that the oscillator dynamics cannot be simply reduced to an
effective temperature in this regime, pointing to a very
significant role of the dynamical effects.

In the adiabatic approximation, the average kinetic energy of the
oscillator has an important role. It describes the effect of the
 \`{}back-action\'{} of the nonequilibrium electronic bath on the
oscillator dynamics and can be used, as shown below, as a tool to
assess the validity of the adiabatic approximation. We show in
Fig.$\ref{Ecin111}$ the behaviour of the kinetic energy $\langle
E_{Kin}\rangle$ for different interaction strengths $E_{p}$ as
function of the bias voltage. First of all, we note that,
regardless the values of $E_{p}$, for $V_{bias}=0$ all kinetic
energy curves show $\langle E_{kin}\rangle=0$. At equilibrium, we
can say that the oscillator  \`{}thermalizes\'{} to the
temperature of the electronic bath ($T_{el}=0$). We have also
plotted two constant energy lines that specify the range of
validity of our approximation, $E=\hbar \omega_{0}/2 \sim
k_{B}T_{D}/2$ and $E=\hbar \Gamma$. At intermediate bias values,
the curves show a departure from the common linear behaviour
observed in the small bias regime, more evident as the interaction
strength increases. The kinetic energy curves corresponding to
$E_{p}=2.0$ and $E_{p}=3.0$ show an interesting plateau at
intermediate bias where increasing the bias does not produce an
increase of the average kinetic energy. Actually, at $E_{p}=3.0$,
we find even a very slight decrease. We also note, in the same
regime, that the velocity distribution probabilities are not
gaussian.
\begin{figure}
\centering
{\includegraphics[width=8cm,height=9.0cm,angle=-90]{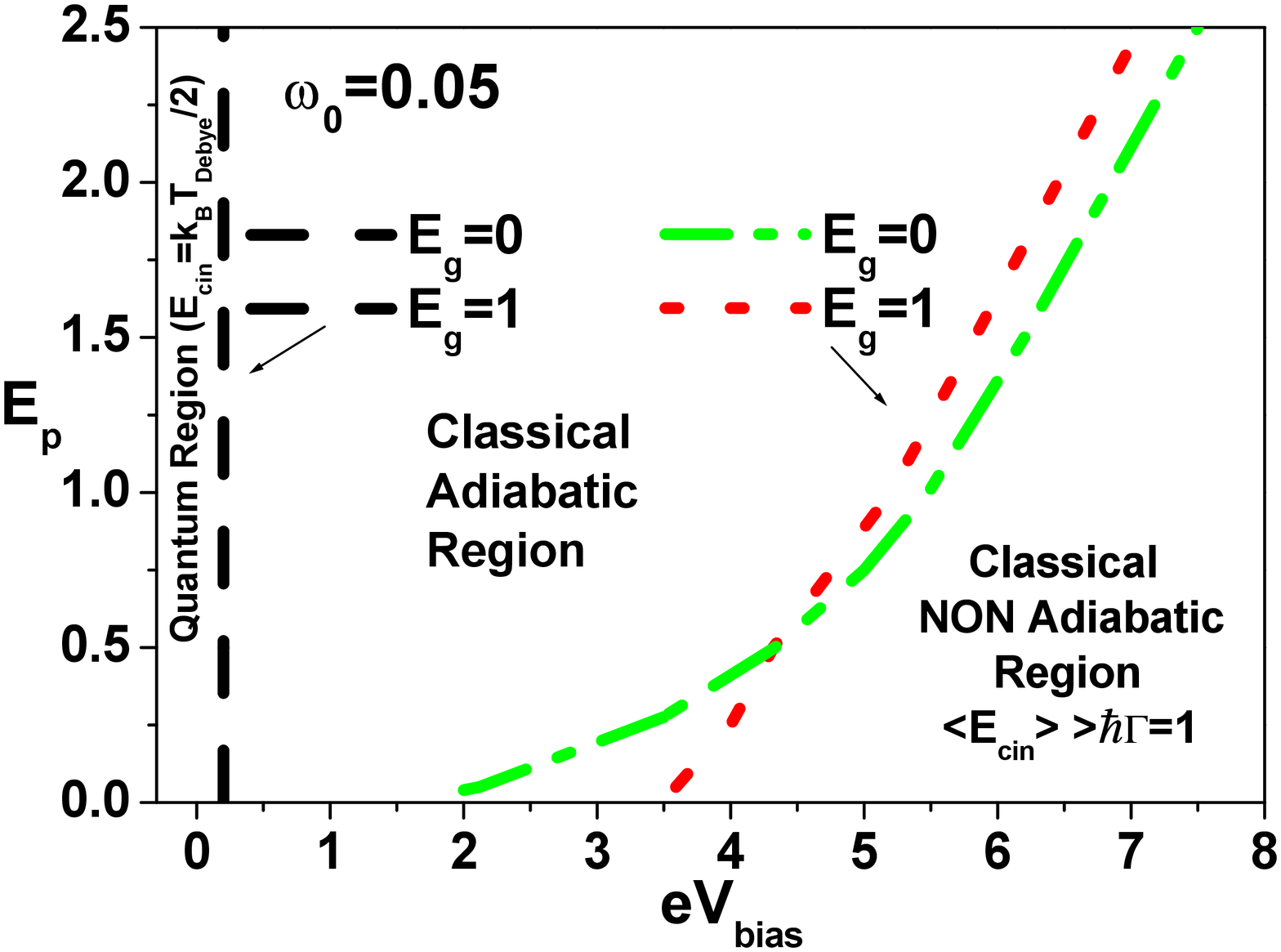}}
\caption{(Color online). Diagram for the range of validity of
classical approximation at fixed adiabatic ratio
$\omega_{0}/\Gamma=0.05$ (the value of $\omega_{0}$ shown in the
figure is in $\Gamma$ units). The dashed (black) line indicates
the QR-CAR crossover for $E_{g}=0$ and $E_{g}=1$. The dotted (red)
and dashed dotted (green) lines indicate the CAR-CNAR crossover
for $E_{g}=0$ and $E_{g}=1$, respectively. $E_{p}$, $E_{g}$ and
$eV_{bias}$ are expressed in unit
$\hbar\Gamma$.}\label{diagfase1sito1}
\end{figure}

\subsubsection{Limits of the adiabatic approach}
\begin{figure}
\centering
{\includegraphics[width=8cm,height=9.0cm,angle=-90]{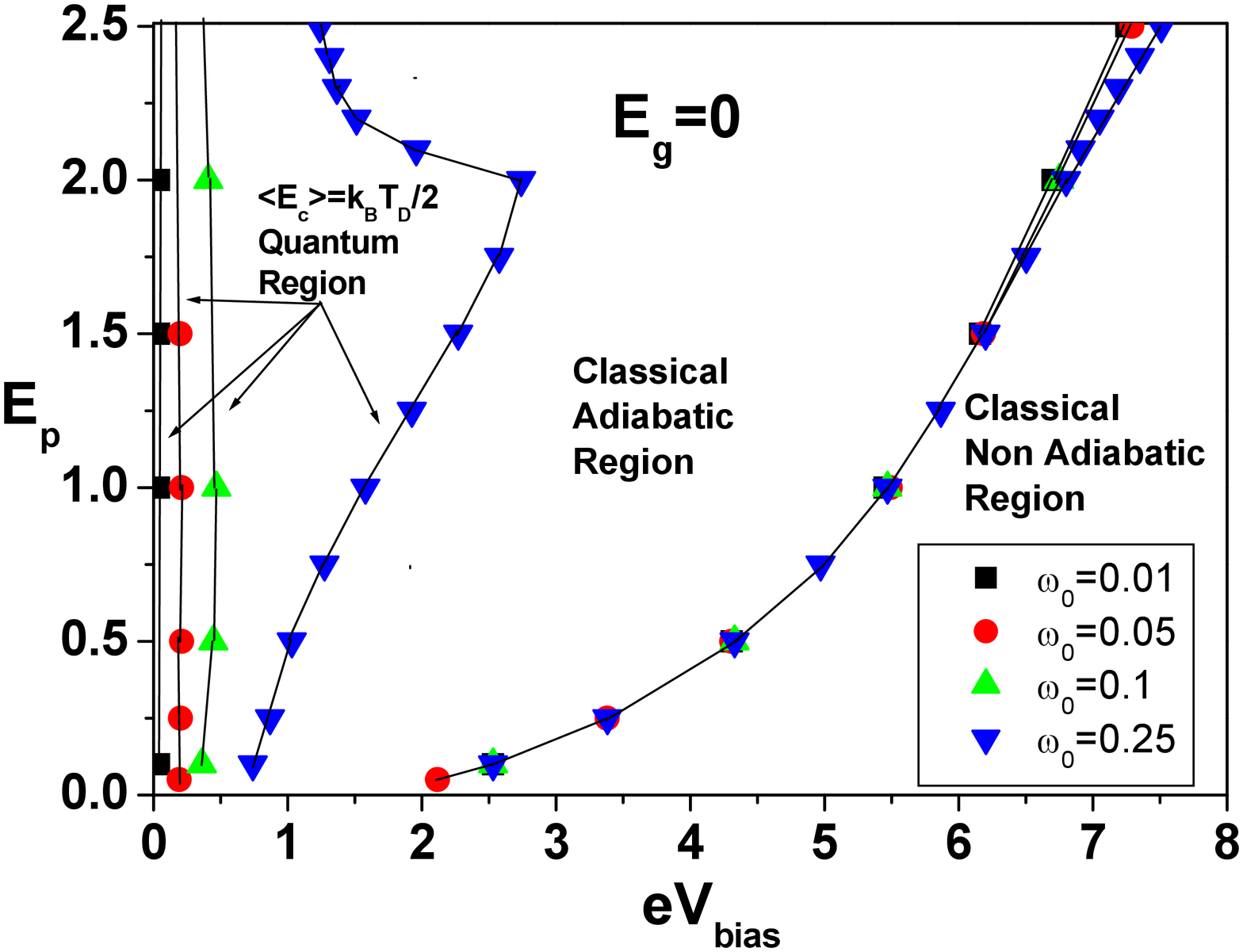}}
\caption{(Color online). Diagram for the validity of classical
approximation at fixed gate voltage $E_{g}=0$ (asymmetric static
potential) for different adiabatic ratios
$\omega_{0}/\Gamma=0.01-0.05-0.1-0.25$ (the values of $\omega_{0}$
shown in the figure are in $\Gamma$ units). $E_{p}$, $E_{g}$ and
$eV_{bias}$ are in $\hbar \Gamma$ units. }\label{diagfase1sito2}
\end{figure}

As mentioned, we can use the average kinetic energy of the
oscillator to fix the range of validity of the adiabatic
approximation. If this energy is lower than the characteristic
Debye temperature of the oscillator $\langle E_{Kin}\rangle<\hbar
\omega_{0}/2 \sim k_{B}T_{D}/2$, we actually explore a region, as
discussed in Ref.$40$, where quantum correlation effects can not
be disregarded. We call this region Non Classical or Quantum
Region (QR). If $k_{B}T_{D}/2 <\langle E_{Kin}\rangle< \hbar
\Gamma$, that is the kinetic energy is lesser than the
characteristic energy scale of the electronic degrees of freedom
and simultaneously greater than characteristic Debye temperature,
a huge number of vibrational quanta (phonons) are excited in the
system. We call this region Classical Adiabatic (CAR). When the
average kinetic energy of the oscillator exceeds the
characteristic energy scale of the electron dynamics $\langle
E_{Kin}\rangle> \hbar \Gamma$, we clearly are going beyond the
limit of adiabatic approximation we start with. We define this
region Classical Non Adiabatic (CNAR). We expect that in the CAR
our approximation is very accurate. By using this data, we are now
able to build up a diagram for the validity of classical
approximation in the plane ($E_{p}$-$V_{bias}$) for different
values of gate voltages (Fig.$\ref{diagfase1sito1}$) and different
adiabatic ratios (Fig.$\ref{diagfase1sito2}$). It is interesting
to note that, in Fig.$\ref{diagfase1sito1}$, the QR-CAR crossover
line is almost independent from the gate voltage in the limit of
small adiabatic ratio. Instead, the CAR-CNAR crossover line is
slightly dependent from the gate voltage showing an enlargement of
the CAR with $E_{g}$. Globally we note that, apart for the QR
(small bias), the CAR enlarges as one increases the electron
oscillator coupling.

As expected, as we increase the adiabatic ratio, the QR expands
reaching great values of bias voltage $V_{bias}$,
Fig.$\ref{diagfase1sito2}$. In particular, observing the
down-triangle (blue) curve (adiabatic ratio
$\omega_{0}/\Gamma=0.25$), one notes that the QR-CAR crossover
line reaches a \`{}\`{}maximum\'{}\'{} in correspondence of
$E_{p}\simeq2$ and $V_{bias}\simeq2.8$. For $E_{p}<2$, the bias
values identifying the QR-CAR crossover increase as the voltage
increases. For couplings $E_{p}>2$, we note an inversion of this
behaviour: the CAR starts to extend for a very large area of the
diagram except for a narrow region at small bias (QR) and for a
region at bigger bias values (CNAR). This means that, even for
intermediate adiabatic ratios, we need sufficiently strong
couplings $E_{p}$ in order to obtain a predominant CAR in the
validity diagram. Moreover, this is due to the fact that the node
between kinetic energy curves and the Debye line occurs in the non
monotonic intermediate bias region (see Fig.$\ref{Ecin111}$). On
the other hand, the CAR-CNAR crossover line is almost independent
from the adiabatic ratio (for not too large interaction strength).
This is what we expect from physical grounds and constitutes a
self-consistent check of our approximation.

\subsubsection{Electronic transport properties}
\begin{figure}
\centering
{\includegraphics[width=8cm,height=9.0cm,angle=-90]{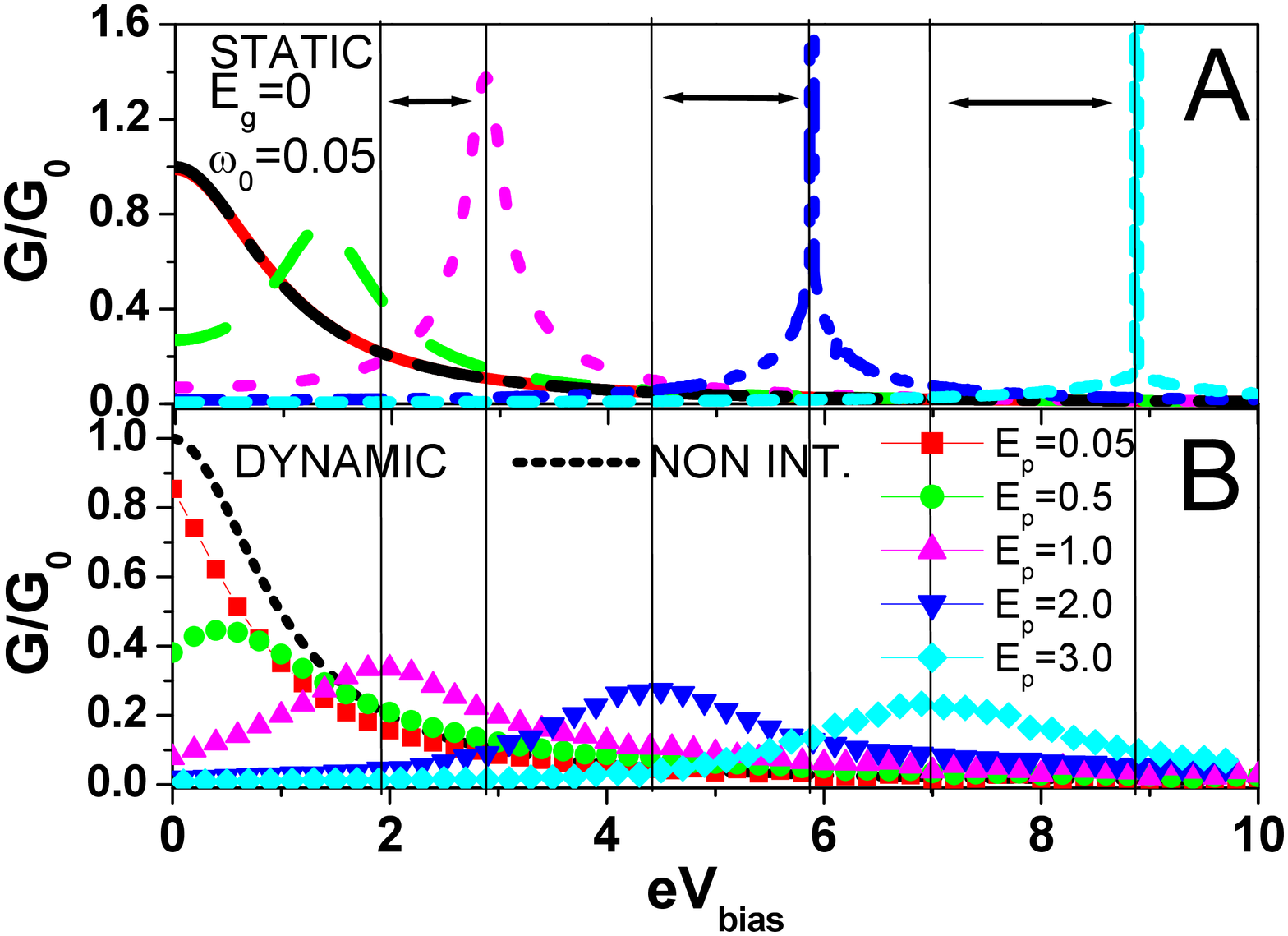}}
\caption{(Color online). Panel A: Conductance (in units
$G_{0}={e^{2}\over 2h}$) in the static approximation as function
of bias voltages, for $\omega_{0}/\Gamma=0.05$, $E_{g}=0$ and
different interaction strengths $E_{p}=0.05,0.5,1.0,2.0,3.0$.
Panel B: Dynamical correction to the conductance for the same
parameter values of panel A. The value of $\omega_{0}$ shown in
the figure is in $\Gamma$ units while all other quantities
($E_{g}$, $E_{p}$ and $eV_{bias}$) are expressed in $\hbar\Gamma$
units.}\label{fispro11}
\end{figure}

\begin{figure}
\centering
{\includegraphics[width=8cm,height=9.0cm,angle=-90]{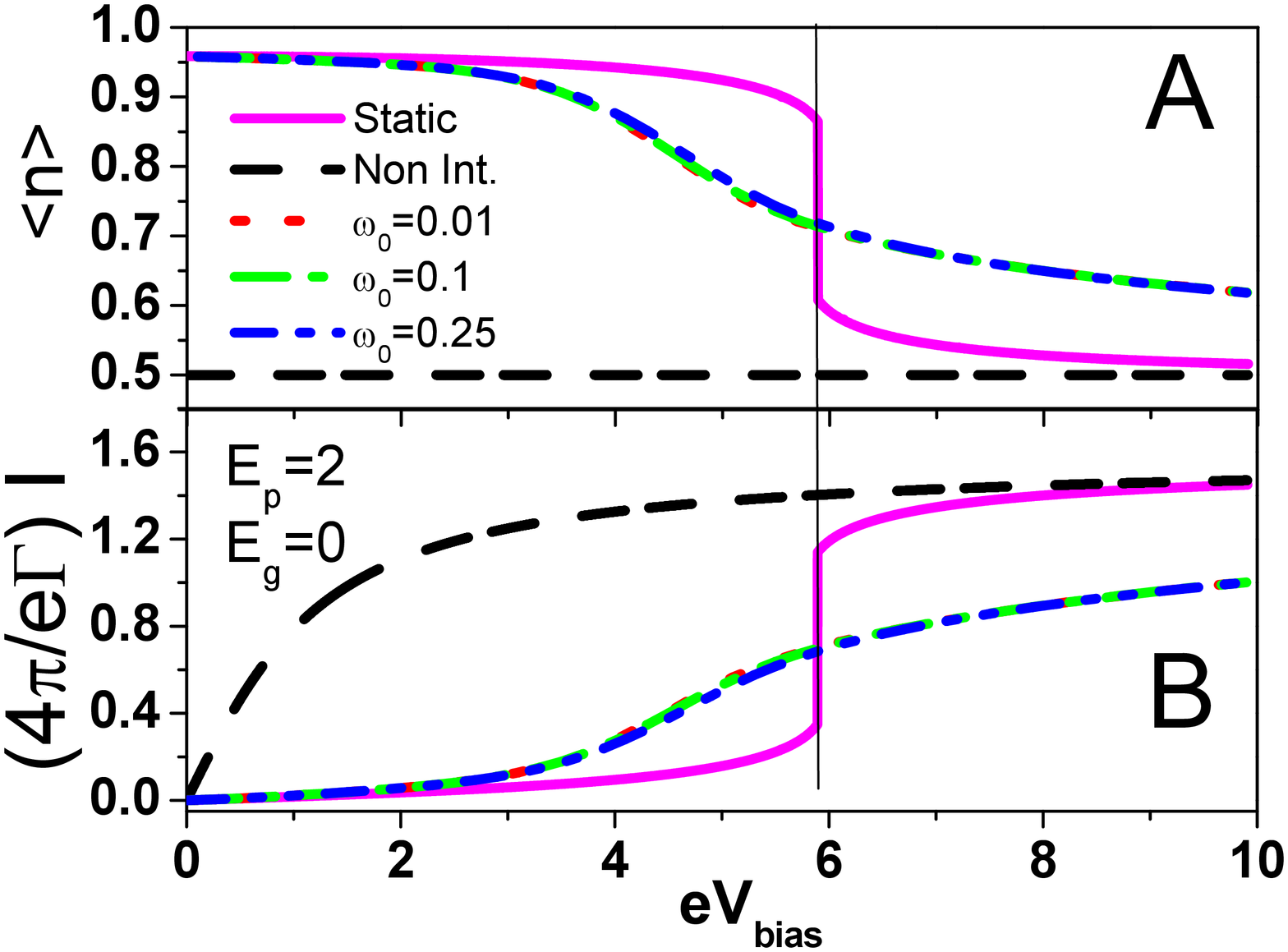}}
\caption{(Color online). Panel A: Electronic occupation as
function of bias voltages, for $E_{g}=0$, $E_{p}=2$ and different
adiabatic ratios $\omega_{0}/\Gamma=0.05,0.25,0.5,1.0$. Panel B:
Current voltage characteristic for the same value of the
parameters of panel A. The values of $\omega_{0}$ shown in the
figure are in $\Gamma$ units while all other quantities ($E_{g}$,
$E_{p}$ and $eV_{bias}$) are expressed in $\hbar\Gamma$
units.}\label{fispro33}
\end{figure}

\begin{figure}
\centering
{\includegraphics[width=8cm,height=9.0cm,angle=-90]{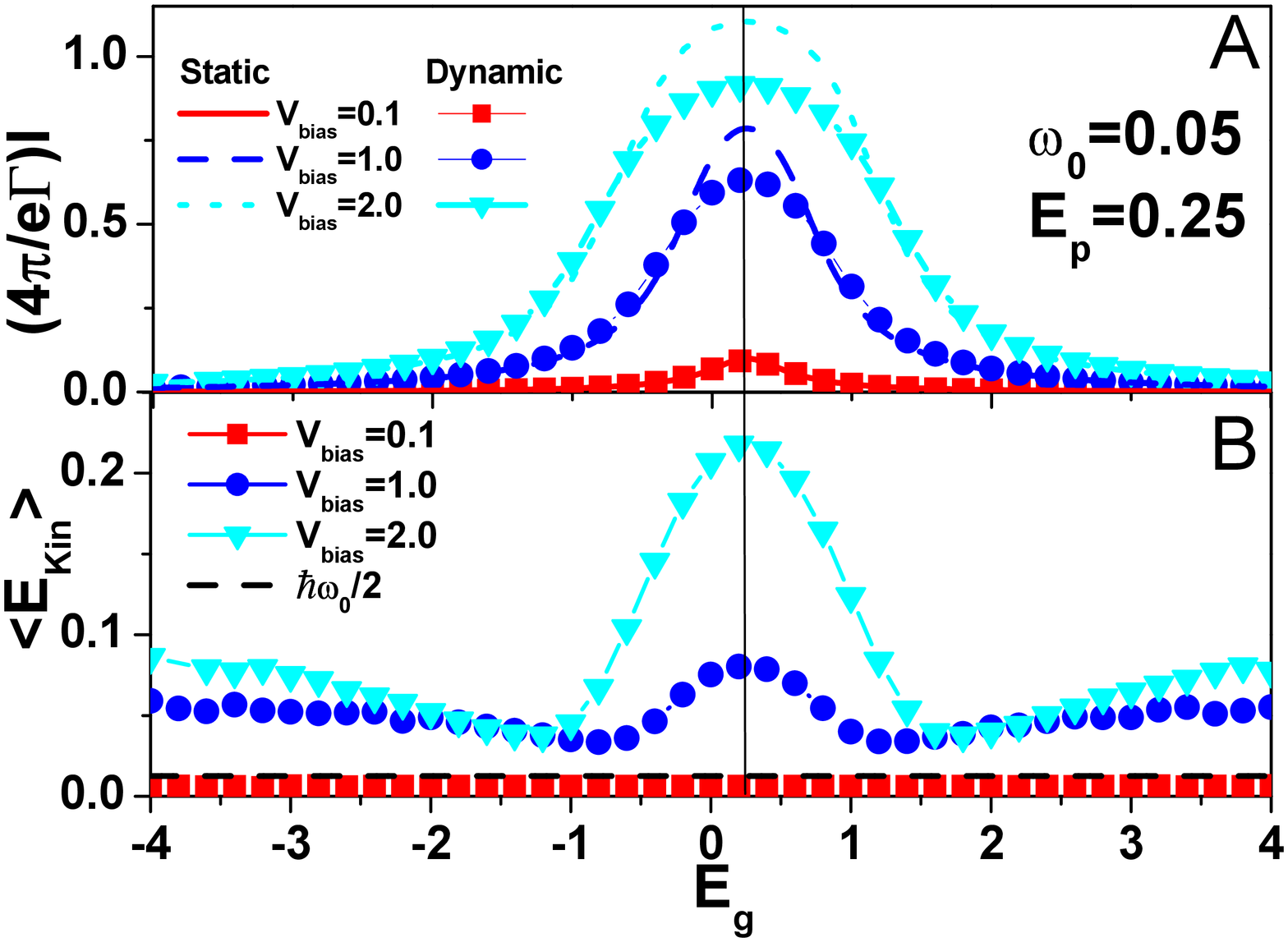}}
\caption{(Color online). Panel A: Current as function of gate
voltages, for $\omega_{0}/\Gamma=0.05$, $E_{p}=0.25$ and different
bias voltages $V_{bias}=0.1,1.0,2.0$. Panel B: Plot of average
kinetic energy as function of the gate for $V_{bias}=0.1,1.0,2.0$.
The value of $\omega_{0}$ shown in the figure is in $\Gamma$
units, while $E_{g}$ and $E_{p}$ are expressed in $\hbar\Gamma$
units. $V_{bias}$ is expressed in $\hbar\Gamma/e$ units.
}\label{fispro22}
\end{figure}

We can now analyze the electronic transport properties resulting
from the average over the dynamical fluctuations of the oscillator
motion. We first study the conductance-voltage curves as function
of the EOC strength (Fig.$\ref{fispro11}$), then we show how the
dynamical fluctuations strongly renormalize the infinite mass
approximation results studying the I-V curves for different
adiabatic ratios (Fig.$\ref{fispro33}$). Finally, we investigate
the dependence of the kinetic energy and of the I-V characteristic
as function of gate voltage studying the properties of the
junction as function of an electric field orthogonal to the
source-drain direction (Fig.$\ref{fispro22}$).

In Fig.$\ref{fispro11}$ we show several conductance curves for
different interaction strengths, $E_{p}=0.05-3$, at
$\omega_{0}/\Gamma=0.05$ and $E_{g}=0$. The comparison between
static (panel A) and dynamical (panel B) approximation is very
interesting. The static solution shifts the non interacting
resonance by a quantity proportional to the polaronic energy
$E_{p}$ (panel A). As one can see, this effect strongly reduces
the small bias conductance. The dynamical correction, on the other
hand, reduces the polaronic shift compared to the static curves
and also broadens (as a result of the very broad nonequilibrium
distribution probabilities $P(x)$) the electronic resonance. In
the intermediate bias regime, we note a strong enhancement of the
conduction far from the electronic resonance where a very small
current is observed in the static approximation. Moreover,
including the dynamical fluctuations, the reduction of the small
bias conductance is less pronounced. We note also that our
dynamical approximation is close to the static solution in the
small bias regime, while is substantially different in
intermediate one. The dynamical corrections strongly renormalize
the static results even for small adiabatic ratios.

We analyze in Fig.$\ref{fispro33}$ the behaviour of the electronic
occupation (panel A) and current voltage characteristic (panel B)
at strong coupling $E_{p}=2$, for different adiabatic ratios
$\omega_{0}/\Gamma=0.01,0.1,0.25$ and at $E_{g}=0$. In the small
bias regime, as a result of strong electron-oscillator
interaction, the molecular level renormalizes itself far below the
chemical potential of the leads. We note a large difference
between the non interacting occupation value ($\langle\hat
n\rangle\simeq 0.5$) and the interacting one ($\langle\hat
n\rangle\simeq 1$). As one increases the bias voltage, many
charges are pumped out the molecular \`{}dot\'{}. In the large
bias regime the stationary charge quantity in the molecular
\`{}dot\'{} reduces approaching the non interacting value
($\langle\hat n\rangle\simeq 0.5$). The nonequilibrium broadening
of the distribution probabilities $P(x)$, then, induces a strong
reduction of the conduction threshold with respect to the static
solution (solid magenta curve in Fig.$\ref{fispro33}$ panel B). We
note a small variation of physical properties with respect to the
adiabatic ratio at intermediate voltages, in the CAR in
correspondence to non-gaussian regime of the distribution
probabilities.

In many molecular transport experiments, one records the current
or the conductance varying an electric field applied on the
molecule (orthogonal to the source-drain direction) at fixed
source-drain voltage. In the panel A of Fig.$\ref{fispro22}$, it
is shown the current for different bias voltages at moderately
small electron-oscillator coupling ($E_{p}=0.25$) as function of
gate voltage. In this regime we have no bistability in the model
($E_{p}<\hbar\Gamma$). We note that the static and dynamical
approximations agree in the small bias regime (solid and square
lines in panel A of Fig.$\ref{fispro22}$). Increasing the bias
voltage, the dynamical correction becomes more important showing a
suppression of the current for small $E_{g}$. This effect is
caused by the spectral weight broadening due to the average over
the position distribution probabilities $P(x)$. From panel B of
Fig.$\ref{fispro22}$ (square (red) line), we learn that, in the
small bias regime, the kinetic energy is independent of the gate
voltage, while, as the voltage increases, it shows a symmetric
drop with respect to polaronic energy $E_{g}=E_{p}$, corresponding
to the symmetric regime. The I-V curve also shares this symmetry.
This effect can be explained observing that, when the \`{}bare\'{}
molecular level and the renormalized one are both in the bias
window, the energy associated to the electronic current flow is
more efficiently exchanged with the oscillator. When the
electronic resonance is far above or below the chemical potential
of the leads there is a less effective coupling between the
oscillator and the electronic subsystem.

\section{The two-site SSH model}
The first step towards a more realistic description of a molecular
junction is to consider a model Hamiltonian composed by two sites
connected by an internal hopping $t$. In particular dimer
molecules,$\cite{Pasu}$ this hopping can be controlled by a
vibrational mode which assists the electron tunneling through the
two molecular sites. In this case, a guess for the molecular
Hamiltonian is given by
\begin{eqnarray}
{\hat H}^{SSH}_{Mol}= E_{g}({\hat d_{1}}^{\dag}{\hat d_{1}}+{\hat d_{2}}^{\dag}{\hat d_{2}})-t(x)({\hat d_{1}}^{\dag}{\hat d_{2}} +h.c.),\nonumber\\
\end{eqnarray}
where we consider, as in the SSH model, an electron hopping
\begin{equation}
t(x)= t-\alpha x \label{approSSH}
\end{equation}
depending linearly on the lattice displacement $x$ associated with
the intermolecular vibrational mode. The molecular sites have a
common energy $E_{g}$ and are described in terms of creation
(annihilation) operators ${\hat d}^{\dag}_{i}({\hat d}_{i})$ ,
$i=1,2$. The SSH model was indroduced to describe the transport
properties of conducting polymers (e.g.
polyacetylene$\cite{SSH1}$) and the two site case represents the
shortest version of a molecular wire.$\cite{Ness}$ A
generalization of this two site model was proposed in Ref.$32$ for
the study the electron transport of dimer molecules interacting
with a single internal vibrational mode.

Most molecular devices studied experimentally so
far$\cite{Reed,Reichert,Pasu}$ have been weakly coupled to the
leads. This corresponds to the bare tunnel broadening
$\hbar\Gamma$ of molecular electronic levels smaller that the
energy required to excite one oscillator quantum (phonon)
$\hbar\omega_{0}$. In the strong-coupling regime, when the
electron-oscillator interaction energy $E_{p}$ exceeds
$\hbar\omega_{0}$, the physics is governed by the Franck-Condon
effect$\cite{Koch05,Koch06,Cav}$, i.e. the tunneling of an
electron onto the molecule with the simultaneous emission or
absorption of several phonons is more probable than elastic
tunneling. The current as the function of voltage exhibits steps
separated by $\hbar\omega_{0}/e$,$\cite{Flensberg}$ and the
conductance shows phonon sidebands.$\cite{Kaat}$

As in the AH model, we study here the case of slow phonons,
$\omega_{0}<<\Gamma$, coupled to a molecular junction driven by a
finite bias, in particular for $eV_{bias}>\hbar\omega_{0}$. As
further approximation, we consider the dynamics of the vibrational
mode \`{}\`{}classical\'{}\'{}.

The structure of the SSH model is very interesting. The direct
coupling of the electron-oscillator interaction to the
intermolecular hopping $t$ suggests that the role of the dynamical
fluctuations becomes crucial to determine the physical scenario.
The total Hamiltonian is
\begin{equation}\label{HtotSSH}
\hat{\cal H}_{TOT}=\hat{\cal H}_{el-SSH}+H_{osc},
\end{equation}
where
\begin{equation}\label{HSSHel}
\hat{\cal H}_{el-SSH}={\hat H}^{SSH}_{Mol}+ {\hat H}_{Tun} + {\hat
H}_{leads},
\end{equation}
with ${\hat H}_{leads}$ and ${\hat H}_{osc}$ given by
Eq.$(\ref{Hleads})$ and Eq.$(\ref{Hosc})$, respectively. The
tunneling Hamiltonian ${\hat H}_{Tun}$ is given by
\begin{equation}
{\hat H}_{Tun}=\sum_{k,L}(V_{k,L}{\hat c^{\dag}_{k,L}}{\hat
d_{1}}+ h.c.)+\sum_{k,R}(V_{k,R}{\hat c^{\dag}_{k,R}}{\hat d_{2}}+
h.c.),\label{HTun}
\end{equation}
indicating that the left (right) lead is coupled only to the
molecular site $1(2)$. In real space, the molecular Hamiltonian
${\hat H}^{SSH}_{Mol}$ is not diagonal. We therefore perform a
transformation which diagonalizes the molecular isolated problem
\begin{eqnarray}\label{Transform}
{\hat c}^{\dag}_{\gamma_{1}} &=& {{\hat d}^{\dag}_{1} + {\hat d}^{\dag}_{2}\over \sqrt{2}},\nonumber \\
{\hat c}^{\dag}_{\gamma_{2}} &=& {{\hat d}^{\dag}_{1} - {\hat
d}^{\dag}_{2}\over \sqrt{2}},
\end{eqnarray}
\begin{figure}
\centering
{\includegraphics[width=7cm,height=7.0cm,angle=0]{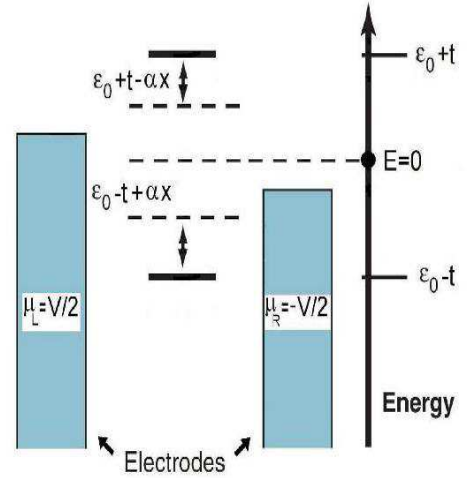}}
\caption{(Color online). Sketch of junction within the SSH model
in an energy scale.}\label{BornOppenheimer2siti}
\end{figure}
with the same transformation for corresponding annihilation
operators. This transformation leaves invariant ${\hat H}_{leads}$
but changes ${\hat H}^{SSH}_{Mol}$ and ${\hat H}_{Tun}$. Explicitly
we have
\begin{eqnarray}
{\hat H}^{SSH}_{Mol} &=& \varepsilon_{\gamma_{1}}(x){\hat
c_{\gamma_{1}}}^{\dag}{\hat c_{\gamma_{1}}}
+\varepsilon_{\gamma_{2}}(x){\hat c_{\gamma_{2}}}^{\dag}{\hat
c_{\gamma_{2}}},\label{HdiagSSH}\\ {\hat
H}_{Tun}&=&\sum_{i=1,2}\Bigg[\sum_{k}({V_{k,L}\over \sqrt{2}}{\hat
c^{\dag}_{k,L}}{\hat c}_{\gamma_{i}}+ h.c.)\nonumber\\
&+&(-1)^{i-1}\sum_{k}({V_{k,R}\over \sqrt{2}}{\hat
c^{\dag}_{k,R}}{\hat c}_{\gamma_{i}}+
h.c.)\Bigg],\nonumber\\\label{HTunSSH}
\end{eqnarray}
where
\begin{eqnarray}
\varepsilon_{\gamma_{1}}(x) &=& \varepsilon + (t -\alpha x),\nonumber\\
\varepsilon_{\gamma_{2}}(x) &=& \varepsilon-(t-\alpha
x).\label{renormlivelliSSH}
\end{eqnarray}
As one can see, the above transformation allows us to take into
account exactly the intermolecular hopping's effect. The molecular
Hamiltonian ${\hat H}^{SSH}_{Mol}$ (Eq.$(\ref{HdiagSSH})$) is
equivalent, at fixed $x$, to that of a non interacting two level
system. In Fig.$\ref{BornOppenheimer2siti}$, a schematic picture
of the junction in an energy representation is shown. We observe
that there are two electronic resonances, corresponding to a
\`{}bonding\'{} and \`{}anti-bonding\'{} states whose position is
renormalized by the electron-oscillator interaction.

From now on we work in the energy space for convenience.

In the following, we will (subsection A) first analyze the coupled
electron-oscillator problem within the SSH model in the limit of
infinite mass for the oscillator. Then, we will construct, as done
in AH model, the stochastic Langevin equation for the dynamics of
the oscillator. In the subsection B we will describe the numerical
results.
\subsection{Out of equilibrium Born-Oppenheimer approximation:
infinite mass (static) case}
 As in AH model, performing the limit $m\mapsto
\infty$, at zero-order static approximation, we neglect the
kinetic energy of the oscillator. The electronic dynamics, with
the oscillator displacement $x$ as a free parameter, is therefore
equivalent in the energy space to a non-interacting two level
problem with energy levels renormalized by the \`{}polaronic\'{}
shift $-\alpha x$, Eq.$(\ref{renormlivelliSSH})$. In what follows,
we consider the case of symmetric coupling of the molecule to the
leads $\hbar\Gamma_{L}=\hbar\Gamma_{R}$ in the wide-band
approximation. Here, we briefly show how to calculate the
generalized potential of the oscillator coupled to the double
\`{}dot\'{} molecular junction.

 Within the Keldysh formalism, we use the equation of
motion approach to calculate the molecular Green functions in
stationary nonequilibrium conditions. In the zero order static
approximation, we have the following equation of motion for the
molecular retarded Green function
\begin{eqnarray}
\left(\begin{array}{cc} \imath\hbar{\partial\over \partial t}- \varepsilon_{\gamma_{1}}(x)+ \imath{\hbar\Gamma_{L}+\hbar\Gamma_{R}\over 4} & \imath{\hbar\Gamma_{L}-\hbar\Gamma_{R}\over 4}\\
\imath{\hbar\Gamma_{L}-\hbar\Gamma_{R}\over 4} & \imath\hbar
{\partial\over
\partial t}- \varepsilon_{\gamma_{2}}(x)+
\imath{\hbar\Gamma_{L}+\hbar\Gamma_{R}\over
4}\end{array}\right)\nonumber
\end{eqnarray}
\begin{equation}
\times\left(\begin{array}{cc}G^{r}_{1,1}(t,t')&G^{r}_{1,2}(t,t')\\G^{r}_{2,1}(t,t')&G^{r}_{2,2}(t,t')\end{array}\right)
=\delta(t-t')\left(\begin{array}{cc}1&0\\0&1\end{array}\right),\label{eqmotoSSH}
\end{equation}
which acquires a $2\times 2$ matrix structure. A similar equation
is valid also for the advanced Green function. In the hypothesis
of symmetric coupling with the leads, one obtains two separate
problems for the molecular energy levels
$\varepsilon_{\gamma_{1}}(x)$ and $\varepsilon_{\gamma_{2}}(x)$,
respectively. The diagonal elements of the retarded Green function
in Fourier space are
\begin{eqnarray}
G^{r}_{i,i}(\omega,x)&=&{1\over\hbar\omega-\varepsilon_{\gamma_{i}}(x)
+\imath\Big({\hbar\Gamma^{L}+
\hbar\Gamma^{R}\over4}\Big)},\;\;\;i=1,2\nonumber\\\label{gritSSHstatico}
\end{eqnarray}
while the non diagonal terms are zero.

The lesser matrix Green function is instead given by
\begin{eqnarray}
G^{<}(\omega,x)=\nonumber
\end{eqnarray}
\begin{equation}
\imath{\hbar\Gamma\over 4}\left(\begin{array}{cc} (n_{L}+n_{R}) |G^{r}_{1,1}|^{2}&(n_{L}-n_{R}) G^{r}_{1,1}G^{a}_{2,2}\\
(n_{L}-n_{R})G^{r}_{2,2}G^{a}_{1,1}
&(n_{L}+n_{R})|G^{r}_{2,2}|^{2}\end{array}\right)\nonumber\label{GlesSSH}
\end{equation}
where, for sake of simplicity, we have dropped the frequency
$\omega$ and the displacement $x$ dependence. The diagonal terms
of the lesser Green function are directly related to the electron
\`{}\`{}densities\'{}\'{} (these obviously not correspond to the
densities in real space)
\begin{equation}
\langle \hat n_{\gamma_{i}}\rangle(x) = {1\over
2}+{1\over2\pi}\sum_{\alpha=R,L}\arctan\Bigg[{\mu_{\alpha}-\varepsilon_{\gamma_{i}}(x)\over\hbar\Gamma/4}\Bigg],\;\;i=1,2.\nonumber\\\label{densSSH}
\end{equation}
For sake of clarity, we show here that the population in real
space of the left and right molecular sites are expressed in terms
of lesser Green functions (Eq.$(\ref{GlesSSH})$)
\begin{equation}\label{densrealiSSH}
\langle \hat n_{i}\rangle(x) = {1\over 2}\int {d\omega\over
2\pi\imath}\Big(G^{<}_{1,1}+G^{<}_{2,2}+(-1)^{i+1}\big(
G^{<}_{1,2}+G^{<}_{2,1}\big) \Big),
\end{equation}
where $i=1,2$ (site $1$ is the left site, site $2$ the right one).

The force exerted on the oscillator is given by
\begin{equation}\label{forSSH}
F_{SSH}=-kx+\alpha(\langle\hat n_{\gamma_{1}}\rangle-\langle \hat
n_{\gamma_{2}}\rangle)(x).
\end{equation}
Taking care of Eq.($\ref{densSSH}$) and Eq.($\ref{forSSH}$), one
can straightforwardly compute the expression of the generalized
potential in nonequilibrium conditions ($\mu_{R}=-eV_{bias}/2$,
$\mu_{L}=eV_{bias}/2$)
\begin{eqnarray}\label{enpotNESSH}
&&V_{SSH}(x)={1\over
2}kx^{2}-\sum_{\alpha=L,R}\sum_{i=1,2}\Bigg[{\mu_{\alpha}-\varepsilon_{\gamma_{i}}(x)\over
2\pi}\times \nonumber\\
&&\arctan\Big({\mu_{\alpha}-\varepsilon_{\gamma_{i}}(x)\over
\hbar\Gamma/4}\Big) - {\hbar\Gamma\over 16\pi}
\ln[16(\mu_{\alpha}-\varepsilon_{\gamma_{i}}
(x))^2\nonumber\\
&&+(\hbar\Gamma)^2]\Bigg].
\end{eqnarray}
This generalized oscillator potential depends parametrically on
the new electronic energy scale introduced in the problem: the
intermolecular hopping $t$ \`{}\`{}hidden\'{}\'{} in
$\varepsilon_{\gamma_{i}}(x)$, see Eq.($\ref{renormlivelliSSH}$).
Furthermore, it depends on the polaron energy, $E_{p}$,  the gate
voltage $E_{g}$, and the bias $V_{bias}$.

\begin{figure}
\centering
{\includegraphics[width=8cm,height=9.0cm,angle=-90]{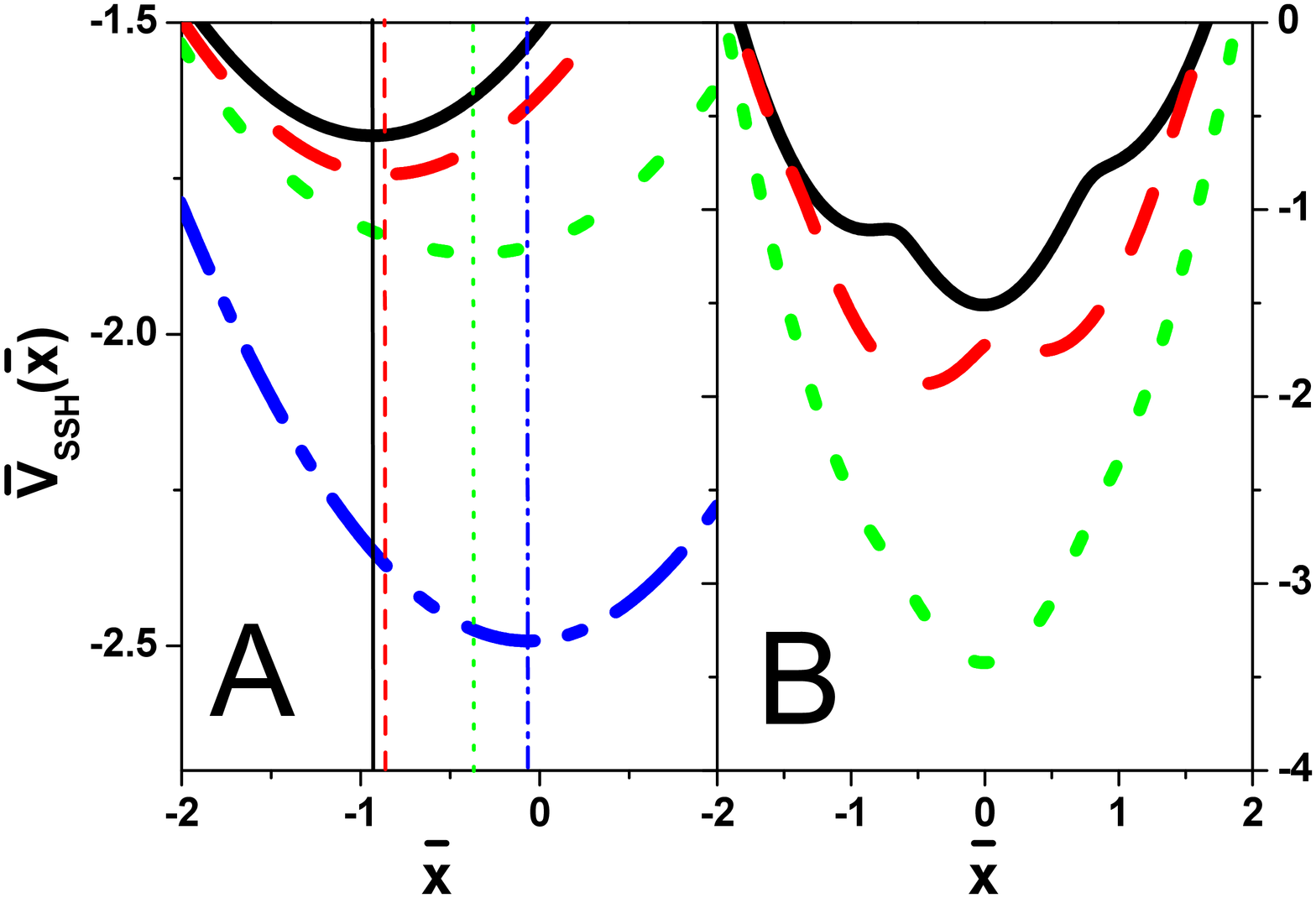}}
\caption{(Color online). Panel A:Spatial dependence of the
dimensionless generalized static potential ${\bar V}_{SSH}({\bar
x})$ at $\omega_{0}/\Gamma=0.1$, $E_{p}=0.2$, gate voltage
$E_{g}=0$, intermolecular hopping $t=2.0$, for different values of
the bias voltage: $V_{bias}=0.0$ (solid (black) curve),
$V_{bias}=3.5$ (dashed (red) curve), $V_{bias}=4.0$ (dotted
(green) curve), $V_{bias}=6.0$ (dashed dotted (blue) curve). The
vertical lines indicate the position of the minima of the
potential. Panel B: Same as above for $\omega_{0}/\Gamma=0.1$,
$E_{p}=1.4$, $t=0.2$, gate voltage $E_{g}=2$ and different values
of the bias voltage: $V_{bias}=0$ (solid (black) curve),
$V_{bias}=4$ (dashed (red) curve), $V_{bias}=8$ (dotted (green)
curve). The potential is expressed in $\hbar\Gamma$ units (${\bar
V}_{SSH}=V_{SSH}/\hbar\Gamma$). $V_{bias}$ values are expressed in
$\hbar\Gamma/e$ units, where $e$ is the electron charge. $E_{g}$,
$E_{p}$ and $t$ are expressed in $\hbar\Gamma$ units. The
dimensionless position variable ${\bar x}$ is defined as ${\bar
x}=x/x_{0}$ with $x_{0}={\lambda\over
m\omega_{0}^{2}}$.}\label{pot_SSH}
\end{figure}

In Fig.$\ref{pot_SSH}$ we present some features of the generalized
potential $V_{SSH}(x)$ which will allow us to understand the
effect of the nonequilibrium electronic system on the
\`{}\`{}static\'{}\'{} stretching of the oscillator (solutions of
the equation $F_{SSH}=0$). Moreover, this will help us to clarify
the role of the dynamical effects in the transport properties that
we will show later.

We focus here on the weak coupling ($E_{p}/\hbar\Gamma<<1$) regime
where moreover the intermolecular hopping $t$ is larger than the
coupling $\hbar\Gamma$ of the molecule with the leads. In the
panel A we show the generalized potential of the SSH model at
fixed EOC strength, $E_{p}=0.2$, intermolecular hopping $t=2.0$,
as function of the bias voltage $V_{bias}$. One can observe that,
as the bias increases, the position of generalized potential
minimum goes from $x\simeq-1$ (corresponding to $\langle
n_{\gamma_{1}}\rangle-\langle n_{\gamma_{2}}\rangle \simeq -1$) to
$x\simeq 0$ (corresponding to $\langle
n_{\gamma_{1}}\rangle-\langle n_{\gamma_{2}}\rangle \simeq 0$).
The oscillator switches from a full stretching configuration
($x\simeq-1$) to a no-stretching one ($x\simeq0$). At equilibrium
(solid (black) curve of Panel A of Fig.$\ref{pot_SSH}$), we have a
physical situation where the renormalized anti-bonding electron
level $\varepsilon_{\gamma_{1}}(x)$ is above the chemical
potential of both leads, while the bonding one
$\varepsilon_{\gamma_{2}}(x)$ is below them. The classical
\`{}\`{}spring\'{}\'{} is fully compressed ($x\simeq-1$) and this
corresponds in real space to molecular sites half-filled ($\langle
n_{1}\rangle\simeq\langle n_{2}\rangle\simeq0.5$). Studying the
electronic populations of left $(1)$ and right $(2)$ molecular
sites (Eq.$(\ref{densrealiSSH})$), one can observe that, if we
increase the bias voltage, the left site starts to empty, while
the right one populates, reaching, for $eV_{bias}^{*}/2\simeq
t-\alpha x(V_{bias}^ {*})$ (hopping value properly renormalized),
a small difference of population roughly equal to $\langle
n_{1}\rangle-\langle n_{2}\rangle\simeq-0.1$. For sufficiently
large bias, the molecular level populations tend again to the
common value $0.5$. As we shall see in next section, the inclusion
of the dynamical effects allows to clarify the physical picture
arising from the above description, in terms of an energy balance
between the electronic and oscillator subsystems.

At static level, it is also interesting to discuss the extremely
strong coupling regime $E_{p}>\hbar\Gamma>t$, for gate voltage
$E_{g}=2.0$ (panel B). In this case, at equilibrium, we are
describing a physical situation where the renormalized bonding and
anti-bonding electron levels are both above the chemical potential
($V_{bias}=0$). The molecular sites in real space are both almost
empty ($\langle n_{1}\rangle\simeq\langle n_{2}\rangle \simeq 0$),
and the oscillator is in a no-stretching configuration $x\simeq0$
(solid (black) curve). Increasing the bias voltage, the
generalized potential develops different minima. At intermediate
bias, one can observe two asymmetric minima near $x\simeq -0.5$
and $x\simeq 0.5$, separated by a potential barrier.  In this
regime, the minimum corresponding to $x\simeq-0.5$ prevails
(dashed (red) curve) and the non-interacting real space
populations $\langle n_{1}\rangle$ and $\langle n_{2}\rangle$ are
asymmetrically distributed between the two sites ($\langle
n_{1}\rangle\simeq0.8$, $\langle n_{2}\rangle\simeq0.2$). Instead,
the interacting real space populations have the same value
$\langle n_{1}\rangle\simeq\langle n_{2}\rangle \simeq 0.25$,
corresponding to a very large current-carrying configuration. In
the large bias regime only the minimum $x\simeq 0$ corresponding
to a small-current-carrying configuration survives. Including the
interaction effects, the left site results almost filled $\langle
n_{1}\rangle\simeq0.9$, while the right one almost empty $\langle
n_{2}\rangle\simeq0.1$, showing that, as result of the strong
electron-phonon interaction, the bias voltage does not manage to
deplete both molecular sites. As we shall see later, the features
of the static potential obtained in this case determine the
possibility to observe in the I-V a strong Negative Differential
Resistance, when the dynamical effects of the oscillator are
neglected.

\subsection{Adiabatic Approximation: calculation of damping and fluctuating term}

\begin{figure}
\centering
{\includegraphics[width=8cm,height=9.0cm,angle=-90]{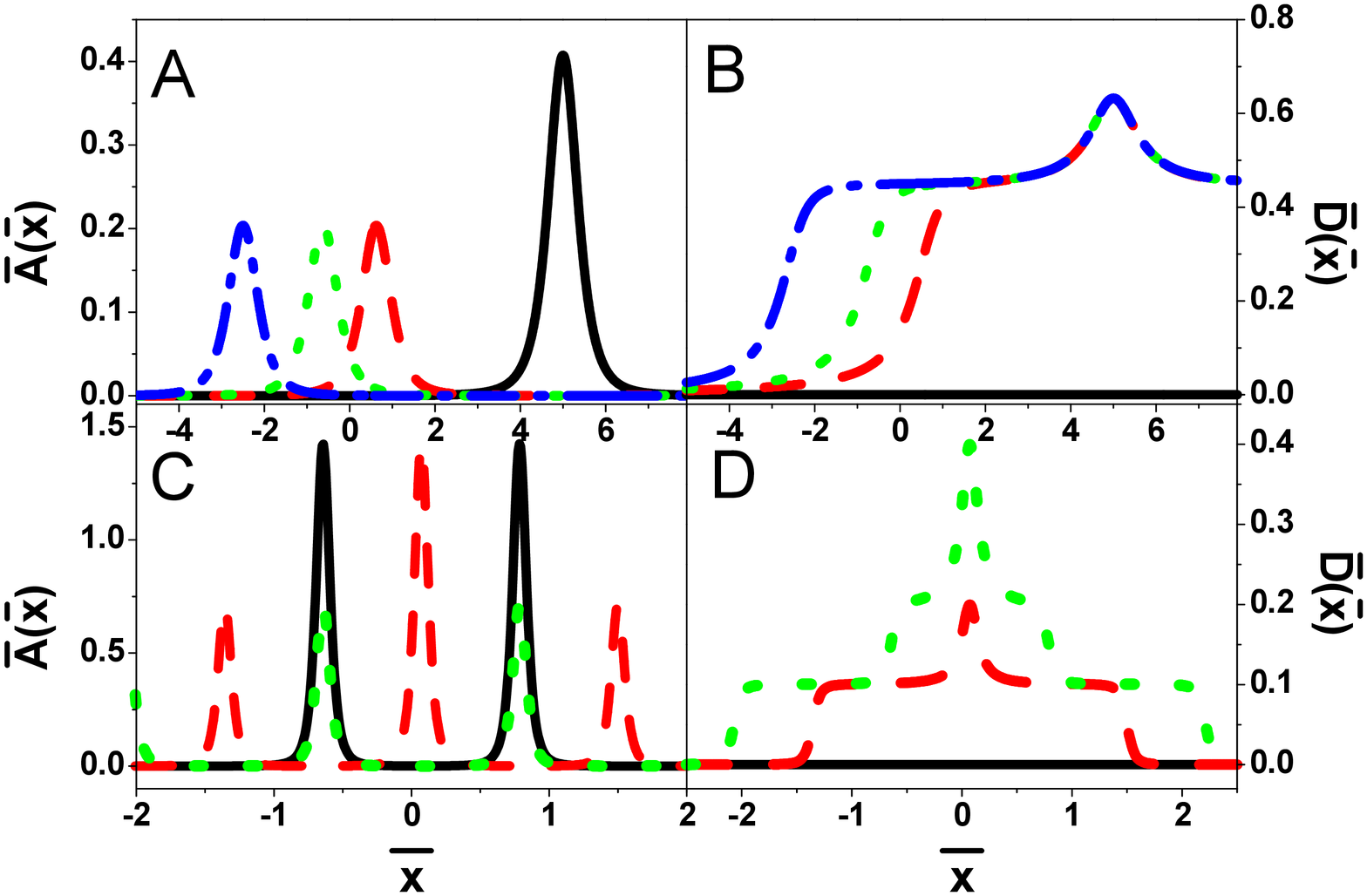}}
\caption{(Color online). Panels A-B: Spatial dependence of the
dimensionless friction coefficient ${\bar A}({\bar x})$ and
fluctuating term ${\bar D}({\bar x})$ at $\omega_{0}/\Gamma=0.1$,
$E_{p}=0.2$, gate voltage $E_{g}=0$, intermolecular hopping
$t=2.0$, for different values of the bias voltage: $V_{bias}=0.0$
(solid (black) curve), $V_{bias}=3.5$ (dashed (red) curve),
$V_{bias}=4.0$ (dotted (green) curve), $V_{bias}=6.0$ (dashed
dotted (blue) curve). Panels C-D: Same as above for
$\omega_{0}/\Gamma=0.1$, $E_{p}=1.4$, $t=0.2$, gate voltage
$E_{g}=2$ and different values of the bias voltage: $V_{bias}=0$
(solid (black) curve), $V_{bias}=4$ (dashed (red) curve),
$V_{bias}=8$ (dotted (green) curve). The friction coefficient is
expressed in $m\omega_{0}$ units (${\bar A}=A/m\omega_{0}$) while
the fluctuating term in $\lambda^{2}/\omega_{0}$ units, (${\bar
D}=D/(\lambda^{2}/\omega_{0})$). $V_{bias}$ values are expressed
in $\hbar\Gamma/e$ units, where $e$ is the electron charge.
$E_{g}$, $E_{p}$ and $t$ are expressed in $\hbar\Gamma$ units. The
dimensionless position variable ${\bar x}$ is defined as ${\bar
x}=x/x_{0}$ with $x_{0}={\lambda\over
m\omega_{0}^{2}}$.}\label{FD_SSH}
\end{figure}

As we have discussed after the Eq.$(\ref{eqmotoSSH})$, the
assumption of symmetric coupling to the leads allows to
disentangle in the energy space the problem for the molecular
bonding and anti-bonding levels $\varepsilon_{\gamma_{i}}(x)$.
Repeating site-by-site the construction introduced in the previous
section for AH model, we can straightforwardly set for our two
site SSH model a Langevin equation for the oscillator dynamics,
very similar to that derived in AH model. The new coefficients,
$F(x)$, $A(x)$ and $D(x)$ are given by
\begin{eqnarray}
F(x) &=& -k x + \lambda
{1\over2\pi}\sum_{\alpha=R,L}\sum_{i=1,2}\arctan\Bigg[{\mu_{\alpha}-\varepsilon_{\gamma_{i}}(x)\over\hbar\Gamma/4}\Bigg],\\
A(x) &=& {16 \hbar k  E_{p}\over\pi \hbar^{2}\Gamma^{2}}
\sum_{\alpha=L,-R}\sum_{i=1,2}\Bigg({1\over
[({\mu_{\alpha}-\varepsilon_{\gamma_{i}}(x)\over
\hbar\Gamma/4})^{2}+1]^{2}}\Bigg), \\
D(x) &=& {k  E_{p}\over \pi}\int d\omega \Big[
G^{<}_{1,1}G^{>}_{1,1} + G^{<}_{2,2}G^{>}_{2,2} +2
G^{<}_{1,2}G^{>}_{2,1}\Big]=\nonumber\\
&=&{2k  E_{p}\over \pi\Gamma}
\sum_{\alpha=L,-R}\sum_{i=1,2}\Bigg\{\Bigg(\arctan({\mu_{\alpha}-\varepsilon_{\gamma_{i}}(x)\over \hbar\Gamma/4})\nonumber\\
&+&{{\mu_{\alpha}-\varepsilon_{\gamma_{i}}(x)\over
\hbar\Gamma/4}\over
[({\mu_{\alpha}-\varepsilon_{\gamma_{i}}(x)\over
\hbar\Gamma/4})^{2}+1]}\Bigg)+ 4\Bigg({1\over
\Big({\varepsilon_{\gamma{1}}(x)-\varepsilon_{\gamma{2}}(x)\over\hbar\Gamma/4}\Big)^{2}+4}\Bigg)\times\nonumber\\
&&\Bigg[\arctan\Big({\mu_{\alpha}-\varepsilon_{\gamma_{i}}(x)\over
\hbar\Gamma/4}\Big)+ {(-1)^{i}\over
\Big({\varepsilon_{\gamma{1}}(x)-\varepsilon_{\gamma{2}}(x)\over
\hbar\Gamma/4}\Big)}\times\nonumber\\
&&\ln\Big(1+\Big({\mu_{\alpha}-\varepsilon_{\gamma_{i}}(x)\over
\hbar\Gamma/4}\Big)^{2}\Big)\Bigg]\Bigg\},\label{dxSSH}
\end{eqnarray}
where in the first line of Eq.$(\ref{dxSSH})$ we have dropped the
frequency $\omega$ and the displacement $x$ dependence in the
Green functions. We end this section briefly discussing some of
the peculiarities of the damping function $A(x)$ and of the
fluctuating term $D(x)$. As regards the damping term (panel A and
C in Fig.$(\ref{FD_SSH})$), one can observe that is located in
suitable points and is strongly space dependent. It is interesting
to note that, as in the AH model case, it survives also for
$V_{bias}=0$ (solid (black) curves in Panels A-C). In this case,
one can also note that, for $E_{p}<<t$ (panel A), $A(x)$ is almost
zero in the interval mostly explored in the dynamics ($-2<x<2$),
while, for $E_{p}>>t$ (panel C), shows two pronounced peaks in
that interval. As one can see, the position of $A(x)$'s maxima is
strongly bias dependent.

As concerns the fluctuating term (Panels B and D in
Fig.$(\ref{FD_SSH})$), one can note that, as in the AH model, it
is identically zero at equilibrium ($V_{bias}=0$). When the bias
increases, it becomes almost different from zero in the region
mostly explored in the dynamics ($-2<x<2$). In Panel B of
Fig.$\ref{FD_SSH}$ one can observe that the spatial extension of
the fluctuating term increases as the bias increases, while, in
Panel D, in the interesting strong coupling regime ($E_{p}>>t$),
it shows a maximum for $x=0$, the no-stretching equilibrium state
of the oscillator. It is important to stress again here that the
space dependence of these terms determines the non-gaussian
character of the distribution probabilities $P(x)$ and $P(v)$ of
the oscillator.

\subsection{Analysis of Numerical results}
As done for the AH model, we here show the results arising from
the numerical simulation of the Langevin equation of the SSH
model. We evaluate the fundamental ingredients of the adiabatic
approximation: the distribution probabilities for the oscillator.
These allow us to calculate the dynamical properties of the
oscillator (average kinetic and potential energy) as well as the
electronic transport properties of the molecular junction.
\subsubsection{Study of the average kinetic energy of the oscillator and limits of the Adiabatic approach}
 First of all, we study the behaviour of velocity distribution
probabilities $P(v)$ resulting from the solution of the Langevin
equation associated to the SSH model. As in AH model, we have
verified that in the small bias regime, regardless the value of
the gate voltages $E_{g}$, the electron-oscillator coupling
$E_{p}$ and the hopping $t$, the \emph{dimensionless} velocity
distribution probabilities $P(v)$ are gaussian. The introduction
of a new energy scale in the problem does not much modify the
physical picture we obtained in AH model in the small bias regime:
the non-equilibrium electronic bath behaves like a conventional
bath for the oscillator with an effective temperature linearly
proportional to the bias voltage. In particular, in the SSH model
case, it is worth noticing that the average kinetic energy
exhibits a slope twice that found in the AH model. This is a
consequence of the transformation Eq.$(\ref{Transform})$ we have
applied on the total Hamiltionian, that renormalized the tunneling
amplitudes with the leads, $V_{k,\alpha}\mapsto
V_{k,\alpha}/\sqrt{2}$. From the physical point of view, we find
that the two electronic \emph{channels} independently contribute
to the oscillator effective temperature, showing that the problem
is equivalent to the \emph{sum} of two single-site junctions.

As we increase the bias voltage, the $(\log(P(v))$ vs. $ v^{2})$
plot starts to deviate from a linear trend, so that, even in SSH
case, the oscillator dynamics cannot be simply reduced to an
effective temperature in the intermediate bias regime.

We also find that, for $0<t/E_{p}<1$ and up to very large values
of the bias voltage, the average kinetic energy shows a behaviour
qualitatively similar to that of AH model (Fig.$\ref{Ecin111}$).
In this regime, we can conclude that the dynamical fluctuations of
the oscillator motion do not \`{}see\'{} the double \`{}dot\'{}
structure of the electronic molecular junction. If $t/E_{p}>>1$,
as we will discuss later, the average kinetic energy shows an
interesting non monotonic behaviour in the intermediate bias
regime (see below, Fig.$\ref{cooling}$).
\begin{figure}
\centering
{\includegraphics[width=8cm,height=9.0cm,angle=-90]{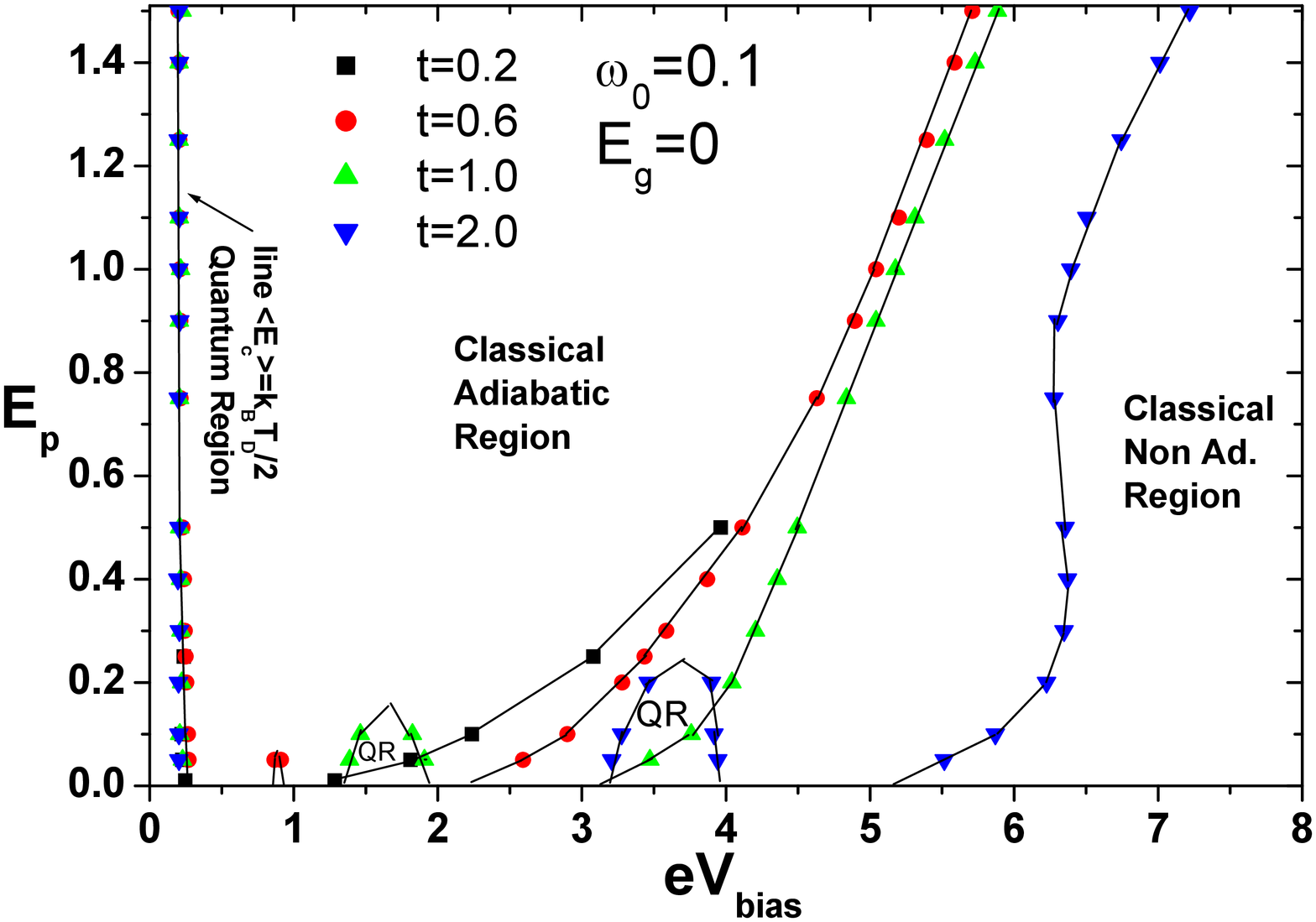}}
\caption{(Color online). Diagram for the validity of classical
approximation at fixed adiabatic ratio $\omega_{0}/\Gamma=0.1$,
$E_{g}=0$, for different values of intermolecular hopping
$t=0.2-0.6-1.0-2.0$. The value of $\omega_{0}$ shown in the figure
is in $\Gamma$ units, while all other quantities ($E_{g}$,
$E_{p}$, $t$ and $eV_{bias}$) are expressed in $\hbar\Gamma$
units.}\label{DiagFaseSSH}
\end{figure}

\begin{figure}
\centering
{\includegraphics[width=8cm,height=9.0cm,angle=-90]{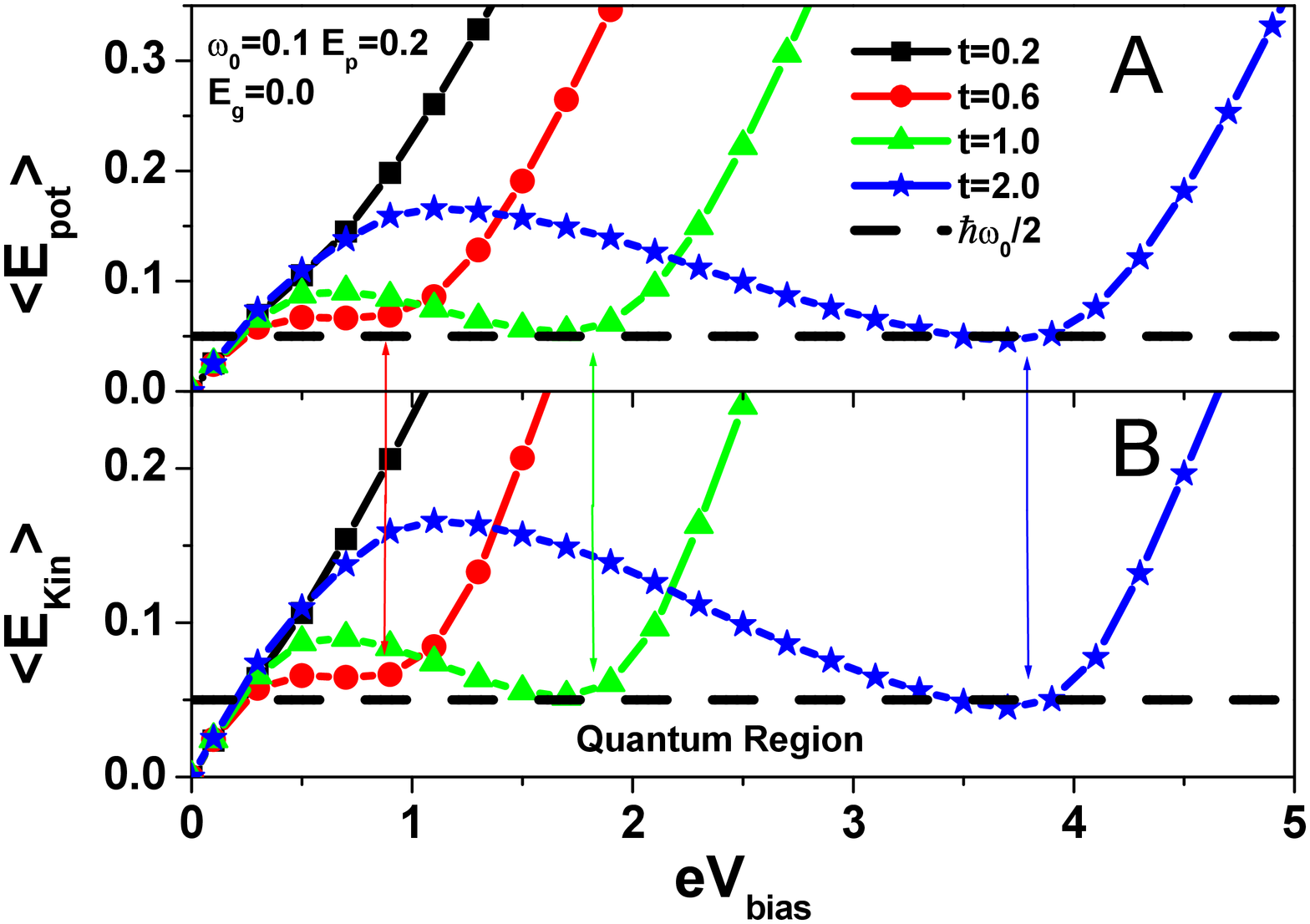}}
\caption{(Color online). Panel A: Average potential energy as
function of the bias for different values of intermolecular
hopping $t=0.2-0.6-1.0-2.0$. Panel B: Average kinetic energy as
function of the bias for values of intermolecular hopping as in
panel A. We note that the introduction of new energy scale makes
the overall energy $\langle E\rangle$ a decreasing function of the
voltage for intermediate values. The value of $\omega_{0}$ shown
in the figure is in $\Gamma$ units, while all other quantities
($E_{g}$, $E_{p}$, $\langle E_{Kin}\rangle$, $\langle
E_{pot}\rangle$, $k_{B}T$, $t$ and $eV_{bias}$) are expressed in
$\hbar\Gamma$ units.}\label{cooling}
\end{figure}

The systematic analysis of the average kinetic energy allows us to
build up a diagram for the validity of classical approximation in
the plane ($E_{p}$-$V_{bias}$), as done for AH model
(Fig.$\ref{DiagFaseSSH}$).  In particular, we study the validity
diagram for different values of intermolecular hopping $t$ and at
fixed adiabatic ratio and gate voltage. In this case, it is
interesting to note that QR-CAR crossover line is almost
independent by the intermolecular hopping in the limit of small
adiabatic ratio. Joining together the results obtained for the AH
validity diagrams (Fig.$\ref{diagfase1sito1}$ and
Fig.$\ref{diagfase1sito2}$), we can conclude that, in the limit of
very small adiabatic ratios, the QR-CAR crossover line is
completely independent by the other energy scales considered in
the problem. The CAR-CNAR crossover line is instead slightly
dependent on $t$ showing the expansion of the CAR. As the
intermolecular hopping $t$ increases, bigger values of bias
voltage are needed to get average kinetic energy values greater
than energy coupling to the leads, $\langle
E_{Kin}\rangle>\hbar\Gamma$. In this case, the intermolecular
hopping $t$ plays the same role as the gate in AH model (see,
Fig.$\ref{diagfase1sito1}$).

A new feature which was not observed in the AH model is the
appearance of small QR for sufficiently small coupling $E_{p}$, at
intermediate bias voltages (Fig.$\ref{DiagFaseSSH}$). For strong
enough electron-phonon coupling $E_{p}$, these regions disappear.
This feature can be understood analyzing the behaviour the average
kinetic energy $\langle E_{Kin}\rangle$ for the parameters
characterizing the QR observed at intermediate bias. As it is
clear form Fig.$\ref{cooling}$, $\langle E_{Kin}\rangle$ can
decrease significantly at intermediate $V_{bias}$. The effect
becomes less and less evident decreasing the intermolecular
hopping and disappears at $t=0.2$. It is interesting to note that
the potential energy curves show almost the same trend (see
Fig.$\ref{cooling}$, panel A). Therefore, for sufficiently large
$t$ and small $E_{p}$, the oscillator overall energy decreases as
a function of bias voltage.

The behaviour of the average energy of the oscillator as function
of bias voltage is determined by net balance of energy exchanged
by the junction: after an increasing trend in the small bias
regime, where the energy pumped by the bias exceeds that ceased to
the electrons by the oscillator, the decreasing behaviour in the
intermediate bias regime is due to the opposite physical
mechanism: the energy ceased to the electron system by the
oscillator exceeds that pumped by the bias.

This \`{}\`{}transition\'{}\'{} occurs for that particular range
of bias voltages where the molecular energy levels are going
through the bias window, with a resulting strong current
enhancement (electronic \emph{resonance}). In particular, when the
electron molecular levels enter the bias window completely, in the
case of symmetric bias unbalance and for $E_{g}=0$, we expect that
the electronic conductance reaches its maximum. Remarkably,
comparing Fig.$\ref{condu3}$ and Fig.$\ref{cooling}$, one can
observe that the conductance maxima correspond to kinetic energy
minima, shifted by a quantity close to the EOC strength $\alpha$.
Physically, as a consequence of the SSH coupling with the
oscillator, the current enhancement is followed by a strong
effective absorption of energy of the electron system from the
oscillator.

\subsubsection{Electronic transport properties}
In order to evaluate the current through the molecular system in
SSH model, we use the Meir-Wingreen formula$\cite{Meir}$ for non
interacting molecular levels, specialized to our two-level case
\begin{equation}\label{Meir}
I(x)={e\over\hbar}\int {d\hbar\omega\over 2\pi}
(f_{L}(\omega)-f_{R}(\omega))Tr\big\{ {\bf
G}^{a}{\bf\Gamma}_{L}{\bf G}^{r}{\bf\Gamma}_{R}\big\},
\end{equation}
where the matrices ${\bf \Gamma}_{L/R}$ are given by
\begin{equation}
{\bf\Gamma}_{L}={\hbar\Gamma\over4}\left(\begin{array}{cc}1&1\\1&1\end{array}\right),\;\;\;
{\bf\Gamma}_{R}={\hbar\Gamma\over4}\left(\begin{array}{cc}1&-1\\-1&1\end{array}\right),\label{GammaSSH}
\end{equation}
and bold ${\bf G}^{r,a}$ indicate retarded (advanced) matrix green
functions (Eq.$\ref{gritSSHstatico}$). We have explicitly
indicated that the current depends on the deformation $x$ of the
oscillator, so that it has to be averaged over the probability
distribution function $P(x)$.

\begin{figure}
\centering
{\includegraphics[width=8cm,height=9.0cm,angle=-90]{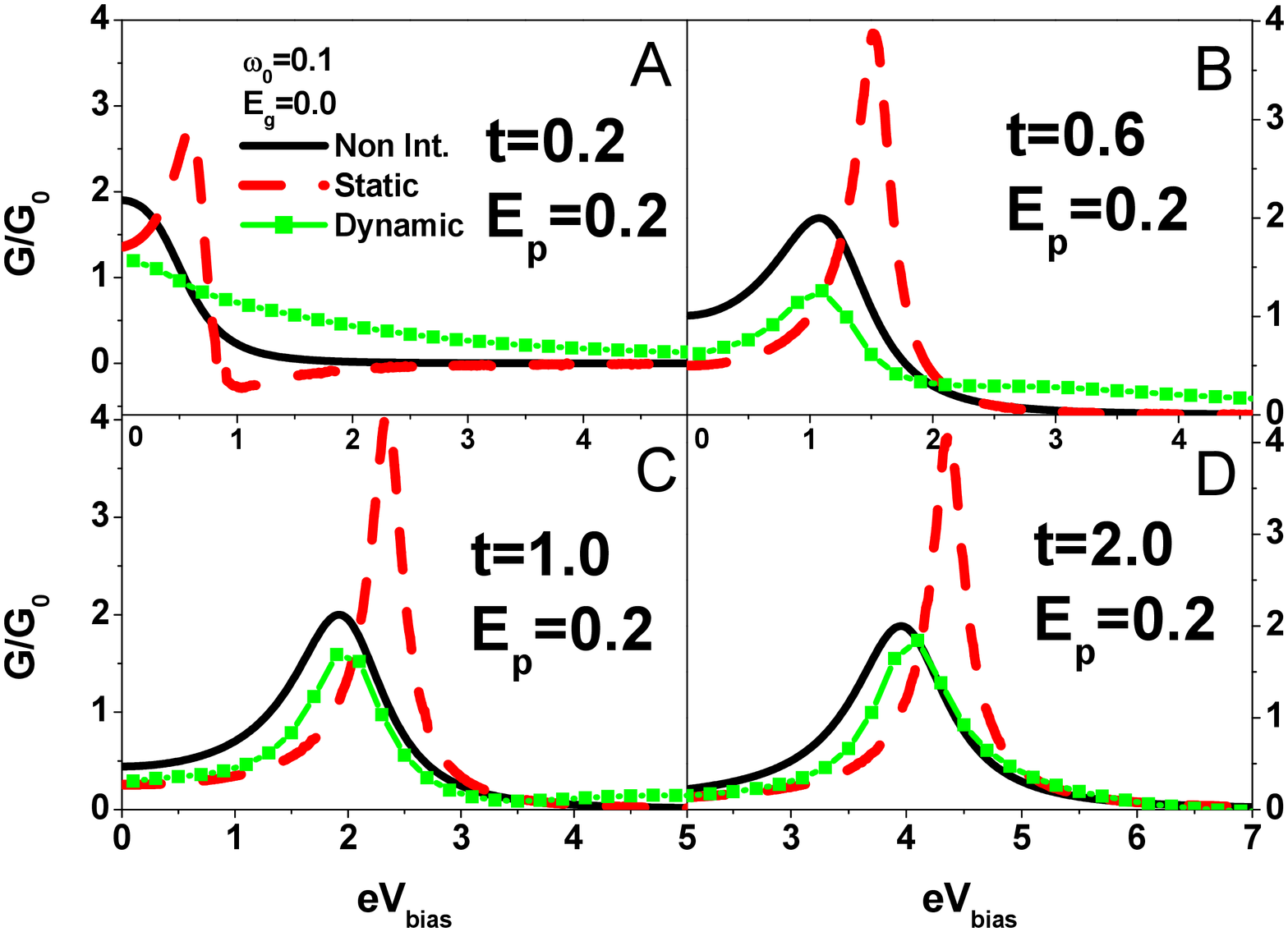}}
\caption{(Color online). Panel A-D: Conductance as function of the
bias at $\omega_{0}/\Gamma=0.1$, $E_{g}=0.0$, interaction strength
$E_{p}=0.2$, for different values of intermolecular hopping
$t=0.2-0.6-1.0-2.0$. The solid (black) line indicates non
interacting curve. The dashed (red) line and the square (green)
lines refer to the static and the dynamical approximation,
respectively. The value of $\omega_{0}$ shown in the figure is in
$\Gamma$ units, while all other quantities ($E_{g}$, $E_{p}$, $t$
and $eV_{bias}$) are expressed in $\hbar\Gamma$
units.}\label{condu3}
\end{figure}

\begin{figure}
\centering
{\includegraphics[width=8cm,height=9.0cm,angle=-90]{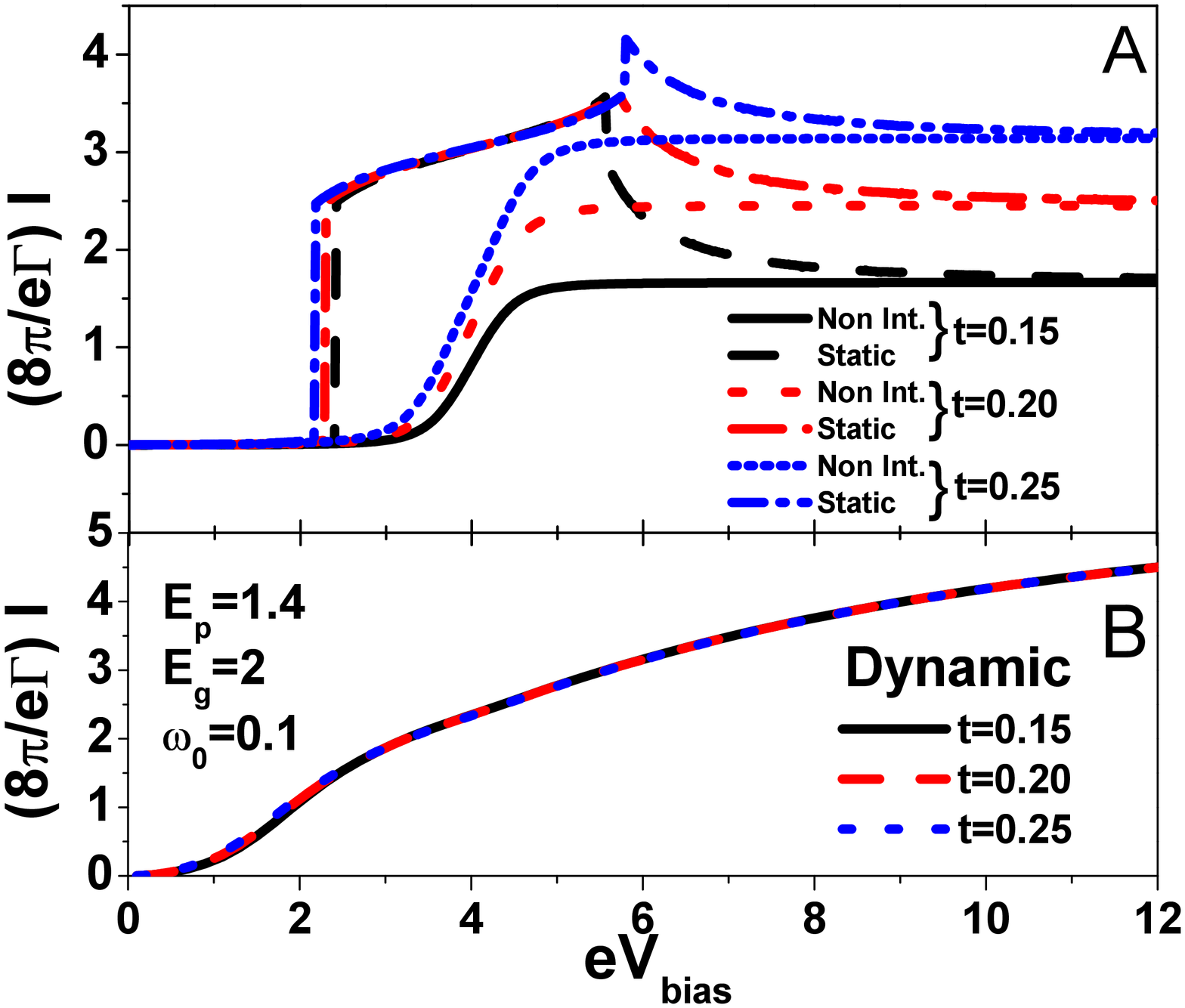}}
\caption{(Color online). Panel A: Current-Voltage characteristic
in the static approximation for $\omega_{0}/\Gamma=0.1$, $E_{g}=2$
and strong coupling $E_{p}=1.4$ for different values of
intermolecular hopping ($t=0.15$ dashed (black), $t=0.2$ dashed
dotted (red), $t=0.25$ short dashed dotted (blue)). The non
interacting quantities ($t=0.15$ solid (black), $t=0.2$ dotted
(red), $t=0.25$ short dotted (blue)) are also shown. Panel B:
Current-Voltage characteristic in the dynamical approximation for
the same parameter values of panel A. The value of $\omega_{0}$
shown in the figure is in $\Gamma$ units, while all other
quantities ($E_{g}$, $E_{p}$, $t$ and $eV_{bias}$) are expressed
in $\hbar\Gamma$ units. }\label{cor2siti}
\end{figure}

Here, we focus on two particular physical regimes, previously
discussed in the analysis of the static approximation: the weak
coupling ($E_{p}<<\hbar\Gamma$) and the strong coupling
($\hbar\Gamma<<E_{p}$) limits, varying arbitrarily the
intermolecular energy scale $t$. As we shall see, in both regimes,
the direct coupling of the electron-oscillator interaction to the
intermolecular hopping makes the role of the dynamical
fluctuations crucial to determine correct results.

In the weak coupling regime, it is interesting to observe that, as
in the AH model, the dynamical corrections renormalize and broaden
the electronic resonances (Fig.$\ref{condu3}$) with respect to the
static solution. In particular, in panel A of Fig.$\ref{condu3}$,
we note that the static approximation exceeds the maximum value of
non interacting conductance and shows a region of negative
conductance at intermediate bias. However, when the dynamical
contributions are included (square (green) curve), the effect on
the conductance is dramatic washing out all the structures
observed in the static approximation. More interesting are the
cases of panels B-C-D of Fig.$\ref{condu3}$, where again the
static approximation shows the spurious result of conductance
greater than $2G_{0}$, while the dynamical approximation
renormalizes and broadens the peak of conductance to bias values
where the static approximation shows small electric conduction.
Even in the weak coupling regime, the inclusion of the dynamical
fluctuations is essential to obtain correct results for the
electronic conduction.

Finally, we examine the electronic transport properties in the
strong coupling regime ($E_{p}>>\hbar\Gamma$), where moreover
$\hbar\Gamma>>t$. In this case, we expect strongly non linear
behaviour of I-Vs in the infinite mass (static) limit for the
oscillator. In Fig.$\ref{cor2siti}$ we show the current voltage
characteristic for strong interaction, $E_{p}=1.4$, at fixed
adiabatic ratio $\omega_{0}/\Gamma=0.1$, gate voltage $E_{g}=2.0$
and for different small values of intermolecular hopping
$t<<\hbar\Gamma$ ($t=0.15$ (black) dashed, $t=0.2$ (red) dashed
dotted, $t=0.25$ (blue) short dashed dotted line). In panel A we
show a comparison between the non interacting and static
approximation. The static approximation shows an interesting
region of Negative Differential Resistance (NDR), as a consequence
of the rich structure of the minima of the generalized potential
described in the previous subsection (see also
Fig.$\ref{pot_SSH}$, Panel B). At intermediate bias voltage, a
strong current currying region appears. This corresponds to
$x\sim\langle n_{\gamma_{1}}\rangle-\langle
n_{\gamma_{2}}\rangle\simeq -0.5$ for which the electronic levels
renormalize in the bias window with an effective energy larger
than the \`{}bare\'{} value. Then, for sufficiently large bias,
the minimum corresponding to $x\simeq 0$ prevails, determining a
strong current reduction due to the drop of the hoppings to their
non interacting \`{}bare\'{} values. As in the case described
above, one can note (panel B) that the dynamical corrections wash
out all the features of the static approximation. There is a very
small conduction threshold after which one not observe NDR
features. Again, we observe that the inclusion of dynamical
corrections are very important for a correct description of the
SSH model while the static approximation can easily lead to
erroneous conclusions.

\section{Conclusions and Discussions}
In this paper we have derived and studied the stochastic Langevin
equation for the dynamics of an oscillator mode coupled to a
voltage-biased molecular junction in the adiabatic limit. Using
the generalization of the Keldysh formalism to time dependent
cases, we were able to show, in agreement with other approaches,
that the oscillator dynamics is controlled by an effective
potential as well as by damping and fluctuating terms coming from
the time depending electronic Green function. Actually, we have
built up an expansion for the molecular level Green function in
the velocity of the oscillator. In this way we have shown that the
quantum effects, \`{}hidden\'{} in the stochastic equation, come
only from the electronic subsystem. Solving numerically the
Langevin equation, we have calculated the position and velocity
distribution probabilities of the classical oscillator. We have
focused our attention on the properties of the velocity
distribution function showing that, for sufficiently large bias
voltage, it loses its gaussian character. In addition, we have
established the range of validity of the adiabatic approximation
underlying the stochastic approach by setting QR (small bias), CAR
(intermediate bias), CNAR (large bias) regimes. The criterion is
based on the comparison of the average kinetic energy of the
oscillator with the Debye temperature ($k_{B}T_{D}\sim \hbar
\omega_{0}$) to distinguish between QR against CAR regimes, and
with electron energy scale ($\sim \hbar\Gamma$) to distinguish
between CAR against CNAR regimes.

We applied our analysis to two simple models of molecules.

For the single site AH model, the analysis of the validity of the
adiabatic approximation has allowed us to build up a diagram for
the validity of classical approximation in the plane
($E_{p}$-$V_{bias}$), showing that the quantum effects are
relevant only in a very narrow region if the adiabatic ratio is
smaller than all other energy scales. Moreover, we have studied
the current-voltage characteristic and the conductance, observing
a dynamical reduction of the polaronic shift and the broadening of
the electronic resonance due to the average on the nonequilibrium
position distribution probability of the oscillator. In the
non-gaussian intermediate bias regime and for sufficiently large
interaction strength, the kinetic energy shows an interesting non
monotonic behaviour. Correspondingly, we observe in the transport
properties a strong enhancement of conduction with respect to the
infinite mass approximation (static limit).

We have also studied the case of a molecular Hamiltonian composed
by a couple of sites interacting with a single vibrational mode in
the SSH model. In this case, because of the direct coupling of the
electron-oscillator interaction to the intermolecular hopping, the
role of the dynamical fluctuations becomes crucial to determine
the physical scenario described by the model. The new
intermolecular electronic hopping energy scale $t$ introduces a
reduction of the CAR in the validity diagram. The new feature is
the occurrence of small QRs for sufficiently small coupling
$E_{p}$, at intermediate bias voltages. For strong enough
electron-phonon coupling $E_{p}$, these regions disappear. In this
region of parameters the average dynamical kinetic energy
decreases as the bias voltage increases. Also the potential energy
curves show this behaviour. Therefore, the oscillator overall
energy decreases as a function of bias voltage. This loss of
energy occurs for that particular range of bias voltages where the
molecular energy levels enter in the bias window. Correspondingly,
as in the AH model case, we observe in the transport properties an
enhancement of conduction with respect to the infinite mass
approximation. Remarkably, in the case of small
electron-oscillator interaction and for $E_{g}=0$, we found that
the maxima of conductance correspond to the minima of the kinetic
energy, shifted by the EOC strength $\alpha$. Finally, within this
model, the dynamical corrections on the transport properties
cancel out completely the \`{}detailed\'{} features (like NDR)
present in the static case. As main result, we can conclude
observing that the inclusion of dynamical effects of the
oscillator motion strongly modifies the physical scenario which
would be obtained by a static description, even if the oscillator
dynamics is much slower than the electron tunneling rate.

We end this section noting that it could be of outstanding
interest to study the possibility to include the quantum
correction to the oscillator dynamics in the small bias regime
classified as Quantum Region (QR). In this direction, Millis
\emph{et al.}$\cite{Millis05}$ find in the quasi-equilibrium
regime $E_{p}>>\omega_{0}>>V_{bias}$ a quantum contribution to the
effective temperature of the oscillator in addition to the
diffusive one. At finite mass $m$, nearby the \`{}gaussian
fluctuation\'{} paths involving small excursions (characteristic
frequency $\omega_{0}$) from the minima of the static potential,
quantum tunneling processes become important. The inclusion of the
quantum corrections in our approach, within the minimal models
considered, is under investigation.

\section{ACKNOWLEDGMENTS} A.Nocera acknowledges F. Cavaliere and
Prof. Sassetti for very useful discussions and CNISM for the
financial support.

\begin{appendix}
\section{Abiabatic Approximation}

Here, we show how the adiabatic approximation on the electronic
Green function Eq.$(\ref{grtimedep})$ works.

In order to implement the adiabatic approximation, it is
convenient to write the Dyson equation for the molecular retarded
Green function (Eq.$(\ref{grtimedep})$)
\begin{eqnarray}\label{DysonT}
G^{r}(t,t')&=&g^{r}(t,t')+\int dt_{1}\int
dt_{2}G^{r}(t,t_{1})\nonumber\\
&+&\Sigma^{r}_{el-ph}(t_{1},t_{2})g^{r}(t_{2},t'),
\end{eqnarray}
where the Green function $g^{r}(t,t')$ already takes into account
the coupling with the leads,
\begin{equation}
g^{r}(t,t')=-{\imath\over\hbar} \theta(t-t')
e^{-\imath({\varepsilon_{0}\over \hbar}-\imath\Gamma/2)(t-t')}
\label{gr0leads}.
\end{equation}
Now, we reparametrize the retarded electron-oscillator self-energy
separating slow and fast times scales (in the following, for sake
of simplicity, we drop the label $_{el-ph}$ of the self-energy)
\begin{equation}\label{self}
\Sigma^{r}(t_{1},t_{2})\mapsto \Sigma^{r}\Big({t_{1}+t_{2}\over
2},t_{1}-t_{2}\Big).
\end{equation}
According to the approach used in Ref.$26$, we expand
Eq.$(\ref{self})$ with respect to the slow mean time
${t_{1}+t_{2}\over 2}$ about a generic time $t_{0}$ belonging to
the interval $[t,t']$
\begin{eqnarray}\label{selfexpansion}
&&\Sigma^{r}\Big({t_{1}+t_{2}\over 2},t_{1}-t_{2}\Big)\simeq
\Sigma^{r}_{0}(t_{0},t_{1}-t_{2})+\nonumber\\
&+&\Sigma^{r}_{1}(t_{0},{t_{1}+t_{2}\over 2},t_{1}-t_{2}),
\end{eqnarray}
with
\begin{eqnarray}
&&\Sigma^{r}_{0}(t_{0},t_{1}-t_{2})=\lambda x(t_{0})\delta(t_{1}-t_{2}) \label{Sig0}\\
&&\Sigma^{r}_{1}(t_{0},{t_{1}+t_{2}\over 2},t_{1}-t_{2}) =
\Big({t_{1}+t_{2}\over 2}-t_{0}\Big)\lambda {\dot
  x}(t_{0})\delta(t_{1}-t_{2}).\nonumber\\\label{Sig1}
\end{eqnarray}
The adiabatic expansion
\begin{equation}
G^{r}(t,t')\simeq G^{r}_{0}(t_{0},t-t')+G^{r}_{1}(t_{0},t-t')
\label{AdiExpansion}
\end{equation}
for the Green function follows from that for the self-energy via
the Dyson equation Eq.$(\ref{DysonT})$. We can now introduce the
Fourier transforms $G^{r}_{0/1}(t_{0},\omega)=\int
d(t-t')e^{\imath\omega(t-t')/\hbar}G^{r}_{0/1}(t_{0},t-t')$. Since
our goal is an adiabatic expansion of the electronic observables
at time $t$, we choose $t_{0}=t$. One can easily show that this is
the only choice able to recover the fluctuation-dissipation
theorem at vanishing bias voltage (equilibrium condition) for the
Langevin equation (Eq.$\ref{Langevin1}$) we derive in Section
$II.B.1$. From the Dyson equation Eq.$(\ref{DysonT})$, taking into
account Eq.$(\ref{Sig0})$ and Eq.$(\ref{Sig1})$, we thus find
\begin{equation}
G^{r}(\omega,t)\simeq G^{r}_{0}(\omega,t)+G^{r}_{1}(\omega,t)
\label{AdiExpansion1}
\end{equation}
with
\begin{eqnarray}
G^{r}_{0}(t,\omega)&=& {1\over \hbar\omega -E_{g}(t)+\imath\hbar\Gamma/2},\label{G0adi}\\
G^{r}_{1}(t,\omega)&=& \imath\hbar{\partial E_{g}\over\partial
t}{\partial
G^{r}_{0}(t,\omega)\over\partial\hbar\omega}G^{r}_{0}(t,\omega),\label{G1adi}
\end{eqnarray}
obtaining a correction which is linear in the velocity of the
oscillator ${\partial E_{g}\over\partial t}=\lambda{\partial
x\over \partial t}$.

\end{appendix}

%\bibitem{Buttiker} R. Landauer, IBM J. Res. Dev. \textbf{1}, 223 (1957); Phil. Mag.\textbf{21}, 863 (1970).

\end{document}